\documentclass[11pt]{article}              

\usepackage[margin=25mm]{geometry} 
\usepackage{graphicx}
\usepackage{float}
\usepackage{colortbl}
\usepackage{subfig}
\usepackage{fancyhdr}
\usepackage{amssymb,amsfonts,amsmath,amsthm}
\usepackage{natbib}
\usepackage{bstnotations}
\usepackage{bm}
\usepackage{soul} 
\usepackage{placeins}
\usepackage{lineno}
\usepackage{xcolor}

\usepackage{authblk}
	
\graphicspath{{./},{./Figures/}}

\newcommand{\Mmean}{\mu_{\tilde{\mathcal{M}}}(\ve{x})}
\newcommand{\Mvar}{\sigma_{\tilde{\mathcal{M}}}(\ve{x})}
\newcommand{\Gmean}{\mu_{\tilde{g}}(\ve{x})}
\newcommand{\Gvar}{\sigma_{\tilde{g}}(\ve{x})}
\newcommand{\eps}{2\Gvar}

\begin{document}
\title{A Two-Level Kriging-Based Approach with Active Learning for Solving Time-Variant Risk Optimization Problems} 

\author[1]{H.M. Kroetz} \author[2]{M. Moustapha} \author[1]{A.T. Beck} \author[2]{B. Sudret}

\affil[1]{Structural Engineering Department, University of S\~ao Paulo, Av. Trabalhador S\~ao-Carlense, 400, 13566-590 S\~ao Carlos, SP, Brazil}

\affil[2]{Chair of Risk, Safety and Uncertainty Quantification,	ETH Zurich, Stefano-Franscini-Platz 5, 8093 Zurich, Switzerland}
\date{}
\maketitle
\abstract{Several methods have been proposed in the literature to solve reliability-based optimization problems, where failure probabilities are design constraints. However, few methods address the problem of life-cycle cost or risk optimization, where failure probabilities are part of the objective function. Moreover, few papers in the literature address time-variant reliability problems in life-cycle cost or risk optimization formulations; in particular, because most often computationally expensive Monte Carlo simulation is required. This paper proposes a numerical framework for solving general risk optimization problems involving time-variant reliability analysis. To alleviate the computational burden of Monte Carlo simulation, two adaptive coupled surrogate models are used: the first one to approximate the objective function, and the second one to approximate the quasi-static limit state function. An iterative procedure is implemented for choosing additional support points to increase the accuracy of the surrogate models. Three application problems are used to illustrate the proposed approach. Two examples involve random load and random resistance degradation processes. The third problem is related to load-path dependent failures. This subject had not yet been addressed in the context of risk-based optimization. It is shown herein that accurate solutions are obtained, with extremely limited numbers of objective function and limit state functions calls.    
	
}

\maketitle

\section{Introduction}
\label{sec:Intro}
Different approaches have been proposed in the literature to solve design optimization problems considering structural reliability. In reliability-based design optimization or RBDO \citep{Hilton1960, Frangopol1985, ABeck2012L, HuDU2014}, a deterministic objective function involving material and manufacturing costs is minimized under reliability constraints. This approach is a natural extension of deterministic design optimization, where deterministic constraints are replaced by probabilistic design constraints. 
A different problem is obtained when structural reliability is part of the objective function. In life-cycle cost or risk optimization \citep{Moses1977, Enevoldsen1994, ABeck2012, TORII2019}, the objective function is formulated in terms of total expected costs, which includes expected costs of failure. These, in turn, are given by the product of failure costs by failure probabilities. Risk optimization allows one to find the optimal point of balance between safety and economy in structural designs. Risk optimization also allows different failure modes to compete with each other. 

Comprehensive literature reviews \citep{Beyer2007, Schueller2009, Chateauneuf2010, ABeck2012L} reveal that the RBDO problem has received much more attention than the life-cycle cost or risk optimization problems. Several very efficient methods have been proposed for solving RBDO. In particular, several methods were designed to overcome the nested optimization loops arising from the use of First Order Reliability Method (FORM) for structural reliability evaluation.  In contrast, not much is found in the literature about solving risk optimization problems. Moreover, it is worthwhile to emphasize that the underlying reliability problems are \textit{time-variant}, due to the presence of stochastic loading, strength degradation (corrosion, fatigue), consideration of inspection and maintenance, etc., which adds another level of complexity.

Assessing the reliability of engineering structures under random load processes, and with consideration of resistance degradation, requires time variant reliability formulations. Unfortunately, analytical or semi-analytical solutions of time-variant reliability problems are limited to very specific cases \citep{Melchers2018}. The up-crossing rate solution is limited to scalar loads with Gaussian distribution. The out-crossing rate solution is limited to polyhedral failure domains. Fast probability integration is subject to convergence problems. Load combination solutions are mainly limited to discrete (pulse-like) processes. Time integrated or extreme value solutions neglect resistance degradation, and so on. Hence, most often time-variant reliability problems have to be solved by Monte Carlo simulation. This has a significant impact in computational costs, which makes the outer optimization loop impractical.  Hence, general methods for solving time-variant risk optimization problems shall involve: a) speeding Monte Carlo simulation via dedicated techniques; and/or b) using surrogate models to simplify (approximate) the underlying time-variant reliability problem. 

With respect to the first point, \citet{Gomes2016} proposed a Monte Carlo-based method which involves finding the roots of the limit state function, in the design space, for each sample. \citet{Rashki2014} and \citet{Okasha2016} proposed efficient solutions for risk optimization involving random design variables. These solutions are based on the ranked weighted average simulation of \citet{Rashki2012}. Regarding the second point, \citet{Echard2011} proposed an active learning method, combining Kriging and Monte Carlo simulation, for reliability analysis. A similar approach was employed in RBDO by \citet{Dubourg2011}, \citet{MoustaphaSMO2016}. \citet{Wang2016} presented an equivalent stochastic process transformation approach for solving general time-variant reliability problems. This approach was  employed by \citet{Wang2018} to solve RBDO problems.

Based on the above observations, this paper proposes a general procedure for solving time-variant risk optimization problems, based on adaptive Kriging \citep{Jones1998,Echard2011, SchoebiASCE2017}. The proposed scheme has some similarities with \citet{Wang2018}; however, herein it is applied for solving time-variant risk optimization problems. Moreover, equivalent stochastic process transformation is not employed herein. Instead, stochastic processes are explicitly evaluated as time series, allowing different features of time variant reliability problems to be addressed. This includes load-path dependency, where failure is not characterized by a point in random load space, but by the whole trajectory of the loads. The drawback of this approach is that Monte Carlo simulations are necessary. To alleviate the computational burden, the proposed approach includes construction of two levels of adaptive Kriging surrogate models. The first level approximates the objective function, allowing for the consideration of different cost terms. The second level approximates the limit state function associated to each cost term of the objective function. This two-stage modeling allows different time-variant risk-optimization problems to be addressed, as shown in the examples section. Different expected improvement functions are conveniently employed to select additional support points for each surrogate model. The efficiency and accuracy of the proposed technique are illustrated in typical time-variant reliability problems, which include random loading, random strength degradation, with discrete or continuous random processes, system-reliability and load path-dependency.

\section{Time-Variant Reliability Problem Statement}\label{sec:TVRPS}
\label{sec:Reliab}
In a context where structures degrade in time, or when loads are described as stochastic processes, it may be important to calculate not only instantaneous probabilities of failure, but the probability of a failure occurring within a certain time interval, sometimes referred to as the \emph{cumulative probability of failure} in the literature. Consider a set $ \ve{X}(t,\omega)  $ of $M = p+q$ elements representing the uncertainties of a given problem, where $ X_j(\omega)$, $j = \acc{1,\ldots,p}$ are random variables, typically describing geometric characteristics and material properties, and $ X_k(t,\omega)$, $k = \acc{p+1, \ldots, p+q}$ are random processes. In this notation, $\omega$ is the outcome in the space of outcomes $ \Omega $. Moreover, let $\ve{d}$ be a vector that gathers together all the system's design parameters. It may include parameters describing moments of random variables, in case tolerances on design dimensions are included in the analysis \citep{MoustaphaThesis2016}. A limit state function $ g(\ve{d},t,\ve{X}(t,\omega))$ defines, for a given $\bm{d}$, safe states if it is greater than zero and failure if it is smaller than zero, so that the boundary between desirable and undesirable structure responses is given by the limit state surface of equation $ g(\ve{d},t,\ve{X}(t,\omega))=0 $:
\begin{equation} \label{eq:Domainf}
D_f(\ve{d},t) = \{\ve{d},\ve{X}(t,\omega) :  g(\ve{d},t,\ve{X}(t,\omega)) \leq 0 \} \qquad \text{is the failure domain,}\nonumber
\end{equation}
\begin{equation} \label{eq:Domains}
D_s(\ve{d},t) = \{\ve{d},\ve{X}(t,\omega) :  g(\ve{d},t,\ve{X}(t,\omega)) > 0 \} \qquad \text{is the safe domain.}
\end{equation}
For each limit state of the problem, the instantaneous probability of failure $ P_{f_{i}} $ at a time $t=\tau$ is calculated as:
\begin{equation} \label{eq:Pfdef}
P_{f_{i}}(\ve{d};\tau) = \Prob{g(\ve{d},\tau,\ve{X}(\tau,\omega)) \leq 0} = \int_{D_f(\ve{d},\tau)} f_{\ve{X}}(\ve{x})\text{d}\ve{x},
\end{equation}
where $\Prob{\bullet}$ indicates the probability of the event $\bullet$ and $f_{\ve{X}}$ is the joint probability density function of the random variables $\ve{X}$ for a given configuration $\ve{d}$ at a time $\tau$.

In the problems studied herein, the quantity of interest is the so-called \textit{cumulative probability of failure} $ P_{fc}(t_1,t_2) $ which is defined for a given configuration $ \ve{d} $ as the probability of occurrence of a structural failure within the time interval $ [t_1,t_2] $:
\begin{equation} \label{eq:PfcDef}
P_{f_{c}}(\ve{d};t_1,t_2) = \Prob{\exists \tau \in [t_1,t_2] : g(\ve{d},\tau,\ve{X}(\tau,\omega)) \leq 0}
\end{equation}

Solutions to time-variant reliability problems include (Melchers and Beck, 2018): the out-crossing approach, the time-integrated approach, the fast probability integration, the nested FORM approach, directional simulation, and specific load combination solutions for discrete pulse-like load processes. These solutions are either approximate, or very specific to particular configurations of the problem. The out-crossing approach is analytical only for Gaussian load processes; for general continuous processes, out-crossing rates need to be approximated by a parallel system sensitivity formulation \citep{Andrieu-Renaud2004,Sudret2008d}. The fast probability integration of \citet{Wen1987} leads to FORM-like solutions, which are potentially unstable due to very small conditional failure probabilities. The nested FORM approach of \citet{Madsen90} applies only to linear or polyhedral failure domains. The directional simulation approach \citep{Melchers1992} requires derivation of conditional strength distributions, which are difficult to evaluate. The time-integrated approach involves extreme-value analysis which is valid only for scalar load processes, and which neglects any strength degradation. Moreover, the approaches named above do not address solution of load-path dependent problems. In these problems, failure at a given time $\tau$ does not depend on the specific combination of loads for time $\tau$, but on the whole time-trajectory of all loads up to time $\tau$ (see \citet{Melchers2018} for details). Thus, the only general method for solving time-variant reliability problems is plain, or brute force Monte Carlo Simulation (MCS). But MCS is known to lead to very high computational burden, for problems involving small failure probabilities. Hence, in order to make plain MCS viable in the solution of time-variant optimization problems, this paper proposes extensive use of surrogate modeling.

\subsection{Simulation-based estimation of the cumulative failure probability}\label{sec:MC_Pfc}
The adopted simulation approach basically consists in drawing sample trajectories of the limit state function over the time interval of interest, and then counting the number of such trajectories for which failure occurs within each time step. In order to do so, the random processes involved in the problem must first be discretized, \ie represented by a finite set of correlated random variables \citep{Sudret2000}. In this work, the \emph{expansion optimal linear estimation} (EOLE) method, after \citet{DerKiureghian1993},  is employed. 

Let $X(t,\omega)$ be a scalar Gaussian random process, with mean $ m(t) $, standard deviation $ \sigma(t) $ and autocorrelation coefficient function $ \rho_X(t_1,t_2) $. An arbitrary number of time points $ P $ are selected in the interval $ [0,\mathcal{T}] $, so that $ t_1 = 0 $ and $ t_P = \mathcal{T} $. The EOLE expansion is then given by:
\begin{equation} \label{eq:EOLE}
X(t,\omega) \approx m(t) + \sigma(t) \sum_{i=1}^{r} \dfrac{\xi_i(\omega)}{\sqrt{\lambda_i}}\phi_i^T C_{t,t_i}(t),
\end{equation}
where $ \acc{\xi_i(\omega), i= 1, \ldots, P} $ are independent standard normal variables, $\acc{ \phi_i , \lambda_i , i = 1, \ldots, r}$ are  the eigenvectors and eigenvalues of the correlation matrix $ \ve{C} $ sorted in decreasing order, with $ \ve{C}_{ij}=\rho_X(t_i,t_j), i,j= \acc{1, \ldots, P} $. The order of the expansion is defined by the number of terms $r \leq P$ that are kept after truncating the series. One usually chooses $r$ in such a way that a significant part of the spectrum of $\ve{C}$ is retained, \emph{i.e.} for an $ \varepsilon \ll 1$:
\begin{equation} \label{eq:trace}
r = \min_{k \in [1,..,P] }\Big\{ k, \sum_{i=1}^{k} \lambda_i \geq (1-\varepsilon ) \textrm{~tr~}\ve{C} \Big\}
\end{equation}
where $\textrm{tr~}\ve{C} = \sum_{i=1}^{P} \lambda_i$ is the trace of the correlation matrix.

Once the random processes are discretized, it is possible to sample trajectories of the limit state function itself. Consider the limit state $g(\ve{d},t,\ve{X}(t,\omega))$ for a given $ \ve{d} $ in the time interval $ [0,\mathcal{T}] $. Samples of the random processes  $X_k(t,\omega),\, k=\acc{p+1, \ldots, p+q} $ are built from the EOLE expansions, and the time independent random variables $X_j(\omega),\, j= \acc{1,\ldots, p}$ are sampled once and remain the same throughout the whole trajectory. Let $ G $ be an array of length $ N $, where $ N $ is the number of time instants in which the continuous time is discretized. The values obtained in the simulation are stored in this array, where each position $ i = 1,...,N $  corresponds to a time $ t_i = (i-1)\cdot \Delta t$, where $ \Delta t = \frac{\mathcal{T}}{N-1}$ is the sampling step, considering a uniform discretization. For each time interval $\bra{t_i,\, t_{i+1}}$, a counter $k_{i+1}$ is defined. Every time $ g $ presents the first outcrossing in the interval $\bra{t_i,\, t_{i+1}}$, all the counters $k_n$, with $ n = i+1 \enum  N $ are increased (i.e. all the remaining counters after the outcrossing are increased). A brute Monte Carlo estimation for the cumulative probability of failure until an arbitrary instant $ t_{i} $, i.e.\ $P_{{fc}_{MC}}(0,t_{i})$,  is given by:
\begin{equation} \label{eq:Pfc}
P_{{fc}_{MC}}(0,t_{i})=\dfrac{k_{i}+k_0}{N_{MC}}, 
\end{equation}
where $ k_0 $ counts the number of failures at $ t=0 $.

\section{Risk Optimization in Structural Engineering}\label{sec:RO}
Defining a structural configuration which is safe and cost efficient at the same time is a challenge for the structural designer. Unfortunately, structures will always be associated with a probability of failure. When the optimal structural configuration  is sought, it is important to optimize in such a way that structural safety is not compromised. This is in general a demanding task, since reducing the dimensions of structural elements tends to reduce structural reliability. Many approaches have been proposed in this context, such as Deterministic Design Optimization (DDO) and Reliability Based Design Optimization (RBDO). Although these two formulation are most often addressed in the literature, they are known to be less general than Risk Optimization, where life-cycle costs are considered throughout the useful lifespan of a structure. A broader review and a comparison between the three approaches can be found in \citet{ABeck2012}. Literature review papers \citep{Chateauneuf2010,Valdebenito2010,Lopez2012} show that many shortcuts have been devised for solving nested optimization loops in FORM-based solutions of RBDO problems. For Risk Optimization problems, however, only two recent papers are known to address efficient solution schemes \citep{Gomes2016,TORII2019}. In this work, a novel surrogate assisted strategy for the solution of time-dependent Risk Optimization problems is developed, so that general cases can be addressed.

\subsection{Risk Optimization Literature Review}\label{sec:ROreview}
	
This section aims to describe some important contributions in Risk Optimization, and to emphasize the lack of efficient surrogate-based techniques addressing the specific complexities in each step of this general structural optimization approach.

For a complete representation of a maintenance planning scheme, the structural degradation that materials tend to suffer over time must be modeled. The most important phenomena to be taken into account in this context, are corrosion caused by chemical agents in the case of concrete structures, and rust and fatigue in case of steel structures. \citet{Joanni2008} developed tools for optimizing design and maintenance strategies of aging structural components. They define time-dependent failure models for deteriorating structural components, which are used in inspection and repair strategies considered in the life-cycle optimization objective function. The authors extend these results to optimal repair and retrofit of existing structures in \citet{Streicher2008}, and present a renewal model for cost-benefit optimization including three different maintenance strategies \citep{Rackwitz2009}.

An interesting application was studied in \citet{Holicky2009}, where a parametric life-cycle optimization is performed for the design of road tunnels. The number of escape routes is optimized in a Bayesian network framework, and some insights are given in the quantification of societal and economic consequences. The optimization makes use of societal risk, defined as a function of an expected value of statistical life (SVSL), which is claimed to be an acceptable compensation cost for one fatality. The life time of the structure and the expected number of causalities (dependent on parameters like the number of escape routes in the tunnel) are also considered. In this study, only one limit state and no maintenance or inspection costs are considered.

\citet{Biondini2009} perform optimization considering the corrosion of reinforcement of concrete structures in aggressive environments. The steel corrosion is modeled as a function of time. The problem statement resembles that of RBDO problems: a deterministic objective function regarding weight of concrete and steel, subjected to probability constraints. The “life cycle” meaning is given by the fact that constraint probabilities of failure are regarded as time dependent, and must be satisfied during the whole life time of the structure. The optimization is performed with a gradient-based method, and the reliability constraints are evaluated with Monte Carlo simulation.

\citet{Okasha2009} proposed a procedure for optimizing not only maintenance intervals, but also the choice between different maintenance actions. In a multi-objective optimization context, two performance indicators are taken into account: the system reliability index and a measure of system redundancy. A digital code is generated for each maintenance scenario, including maintenance type, the structural component to which it applies, and a binary digit representing the application or non application of such action. A genetic algorithm is then run to select the most favorable scenario. The time between each intervention is also optimized. The structures initial costs and the costs of failure are not optimized, since material and geometrical parameters of the structures are not considered to be design variables.
 
A broad framework is presented by \citet{Taflanidis2009}, who propose techniques for the estimation and optimization of the life-cycle cost for dissipative devices. The considered life-cycle costs include performance-based earthquake engineering considerations, initial, repair and replacement costs. A simulation approach called Stochastic Subset Optimization is proposed to establish a global sensitivity analysis for the performance using a relatively small number of system analyses. A second step is suggested for the refinement of the solution, utilizing classical optimization algorithms.

In 2010, another comprehensive review concerning structural optimization under uncertainties was published \citep{Valdebenito2010}. Surrogate models are entitled a section in this review, where works with artificial neural networks, support vector machines and Kriging are cited. There is no mention of risk optimization, and all works regard the classic RBDO approach.

Papers without developments in the theory itself, but with life-cycle optimization applications are easily found in the literature. \citet{WenYK2001b} studied a nine-story office building subjected to earthquake and wind loading. The hazards are described by a Poisson process, and the methodology described in \citet{WenYK2001a} is employed. \citet{Saad2016} presented a comprehensive application to reinforced concrete bridges, considering degradation over time and use-associated costs. \citet{LiHu2014} employed metaheuristics to solve a multi objective risk optimization problem. The problem is analyzed in the context of performance based wind engineering. Elements of risk optimization theory are often used in wind turbines engineering, where optimum maintenance strategies are a key element for total costs reduction. 
An overview of RO applications to this subject can be found in \citet{Nielsen2014}.

\citet{Kroetz2019} developed a time-dependent procedure for the reliability assessment of reinforced concrete beams subjected to corrosion. Even with application of efficient modeling techniques, such as the boundary element method, the authors concluded that surrogate models could be necessary for more complex studies. The idea of utilizing surrogate models to aid in optimization problems under uncertainty is not new (e.g. Adaptive Kriging employed in the solution of RBDO problems \citep{Bichon2013,Dubourg2011a}), but only recently surrogate models have been employed for the specific problem of risk optimization. 

In \cite{Gomes2013},  Artificial Neural Networks are trained to represent limit state function associated with structural failures, in a context where the optimization is carried out by a hybrid algorithm. Particle swarm optimization (a zero-order heuristic global optimization algorithm) scans the problem domain, and a gradient based algorithm is used to refine the result. Only initial and expected failure costs are considered. The disregard of other costs, such as inspection and repair costs, seems to be a common practice in stochastic optimization. 
\citet{Aissani2014} explain that most of these costs can be more or less well estimated, but the failure cost is particularly important, because it is difficult to evaluate and greatly affects optimal solutions. Therefore, other costs can be considered constant in some cases, which means that they do not affect the optimization process. In the same work, the idea of considering nonlinear costs of failure is presented. In the proposed model, failure costs increase linearly with the probability of failure up to a certain threshold, from which the costs increase exponentially. Ideally, this accounts for catastrophic failures which imply in further social or environmental damage.

Surrogate models were also employed by \citet{Carreras2016}. A multi-objective problem considering a prototype of a building which is retrofitted through installation of insulation materials is considered. Insulation material and thickness of the insulation layer are optimized, so that the economic and the environmental performance of the building are optimum. An objective reduction strategy is proposed, and cubic spline surrogates are employed to reduce the computational burden of the analysis.  Life cycle optimization targeting for energy efficiency have also been performed with the aid of Support Vector Machines by \citet{Eisenhowe2012}.

A general surrogate assisted stochastic optimization is proposed by \citet{ZhangTafl2017a}. A sequential approximation optimization strategy is adopted, where the initial design problem is decomposed into cycles of Kriging-based, sub-region constrained optimization problems. Adaptive Kriging models are built from an augmented design of experiment (which considers both design and random variables). This idea was extended to multi-objective stochastic optimization problems in \citet{ZhangTafl2017b}.
	
The explicit simulation of limit state equations related to the risk optimization problem throughout the whole time-series of the analysis was not studied in any of the references, possibly because of the difficulties regarding the high computational cost of such procedure. In this work, specific adaptive surrogate models are efficiently built to address both the complexities present in the steps of structural reliability analysis and global optimization.
	
\subsection{Risk Optimization Formulation} It is well known that Risk Optimization is more general than alternative approaches such as deterministic design optimization and reliability based design optimization \citep{ABeck2012}. When the total life-cycle cost of a structure is of interest, a comprehensive approach should account for the expected cost of failure. Hence, risk optimization is employed herein. In this context, the function to be minimized is the total life cost  $ C_T(\ve{d}) $, defined by:

\begin{equation}\label{eq:CT}
C_T(\ve{d}) = C_I(\ve{d}) +  C_O(\ve{d}) + C_{I\&M}(\ve{d}) +C_{EF}(\ve{d}), 
\end{equation}
where $ \ve{d} \in \mathbb{D}$ is a given design configuration. This cost is composed of various terms, namely the \textbf{I}nitial design costs $C_I$, \textbf{O}peration costs $C_{O}$,  \textbf{I}nspection and  \textbf{M}aintenance costs $C_{I\&M}$,  and the \textbf{E}xpected cost of \textbf{F}ailure $C_{EF}$, defined as: 

\begin{equation} \label{eq:Cef}
C_{EF} = \sum_{j=1}^{N_{ls}} P_{f_j}C_{f_j},
\end{equation}
where $ j = \acc{1, \ldots, N_{ls}}$ enumerates different limit states associated with a possible failure that occurs with a probability $ P_{f_j} $ and whose cost is $ C_{f_j} $. Design and reliability constraints can also be considered, so that the optimization problem can be cast as:

\begin{equation}
\begin{split}
&\ve{d}^\ast = \arg \min_{\ve{d} \in \mathbb{D}} C_T(\ve{d}),  \\
\text{subject to:} \quad & P_{f_j} \leq \bar{P}_{f_j}, \quad j = \acc{1 \enum N_{ls}},\\
\end{split} 
\label{eq:bigtotalcost}
\end{equation}
where $\bar{P}_{f_j}$ are target failure probabilities that shall not be exceeded, for each limit state.

Constraints are often unnecessary in this type of problem, since the probabilities of failure are directly defined in the objective functions. Although reliability constraints can be considered in order to comply with standards, the solution domain  $\mathbb{D}$ may also be limited by bound constraints.

Ideally, the four cost terms in Eq. (\ref{eq:CT}) should be addressed simultaneously, as for some structures, the initial costs also depend on inspection and maintenance policies \citep{Gomes2013b, GOMES2014}. For other structures, initial or construction costs have low dependency with inspection and maintenance costs \citep{Lee2004}. Without loss of generality, and to maintain simplicity, the problems addressed in this paper do not consider inspection and maintenance costs. Expected failure costs, however, have strong dependency with initial costs, as construction costs affect both the fixed cost of failure and the failure probability. Expected costs of failure are largely dependent on the use of the structure and the surrounding environment; hence, they are very application dependent. In this paper, for simplicity and without loss of generality, costs of failure are assumed as a multiple of initial costs. Initial or constructions costs vary with the amount and quality of structural materials, as shown in the examples section. Cost of workmanship may also be considered.”

Since the life cycle of a structure may be considered to span over years or decades, the costs to be optimized cannot be directly treated. Economic changes over time would make the considered values unrepresentative. In order to account for this effect, the structural life time can be discretized, and all costs brought to present value considering discount rates over each period (e.g. yearly discount rates). This way, cumulative failure probabilities associated with each given period can be considered to compose the expected cost of failure as follows \citep{Saad2016}:
\begin{equation} \label{eq:Cefpv}
C_{EF}^{PV}({\mathcal{T}}) = \sum_{j=1}^{N_{ls}} \sum_{n=1}^{\mathcal{T}} \dfrac{P_{fc_{jn}}C_{f_{jn}}}{(1+\eta)^n}
\end{equation}
$ C_{EF}^{PV} $ is the expected cost of failure in present value, $P_{fc_{jn}}$ and $C_{f_{jn}}$ are, respectively, the so-called \emph{cumulative probability} and cost of failure of the $j$-th limit-state in year $n$, and $\eta$ is the discount rate, herein adopted as $1\%$ per year. In the remainder of this paper, instead of Eq.~\eqref{eq:Cef}, Eq.~\eqref{eq:Cefpv} will be used to compute the expected cost of failure in Eq.~\eqref{eq:CT}.

\section{Surrogate-aided optimization}
The solution of the problem in Eq.~\eqref{eq:CT} relies on optimization techniques which would usually require thousands of calls to the objective function $C_T$. Furthermore, the evaluation of a single cost $C_T(\ve{d})$ requires to solve a time-variant reliability analysis using Monte Carlo simulation. The associated cost amounts to millions of calls to the limit-state function. Solving naively this problem as introduced above would therefore be extremely time-consuming. This becomes even more intractable when the limit-state function involves expensive-to-evaluate computational models.

To address this challenge, surrogate modeling is used in this paper. The basic idea is to replace a time-consuming black box model by an analytical proxy that can be evaluated millions of times at practically no cost. Several surrogate modeling techniques have been introduced in the literature to solve optimization and reliability analysis problems, \eg response surface models \citep{Foschi2002}, polynomial chaos expansions \citep{Blatman2008, BlatmanPEM2010, BlatmanICASP2011, Kroetz2017a}, support vector machines \citep{BourinetHDR, Deheeger2007}, artificial neural networks \citep{Papadrakakis2002, Kroetz2017a, Lekhy2017} or Kriging \citep{Picheny2008, Viana2009, Dubourg2011, MoustaphaSMO2016, Kroetz2017a}. In this work, we are interested in Kriging as it features a built-in error measure that arises from epistemic uncertainty and which allows for the development of active learning techniques. Such techniques allow one to reduce the computational cost of building the surrogate model by controlling its accuracy only in confined regions of the input space. A brief literature review about Kriging and well-known adaptive bulding strategies is presented in the following section, and the techniques adopted in this work are detailed in Appendix A.
\subsection{Adaptive Kriging in Structural Optimization}

The idea of Kriging dates back to the decade of 1950, where it was first applied for improvement predictions in the context of geostatistics. A historical review of the method is presented in \citet{Cressie1990}. This technique has been used to aid in the solution of structural reliability problems in the works of \citet{Romero2004} and \citet{Kaymaz2005}, and found many different applications since. \citet{Echard2011} proposed an adaptive learning method where the design of experiment from which the Kriging surrogate model is built is updated depending on a discrete evaluation of the interest region in reliability space. Reliability-based design optimization using adaptive Kriging was addressed in the work of \citet{Dubourg2011}. In order to further reduce computational costs, \citet{Echard2013} coupled importance sampling Monte Carlo with Kriging, in solution of structural reliability problems, showing that the technique is suitable even when small probabilities of failure are present. A broad review about different Kriging applications in the context of structural reliability can be found in \citet{GasparPEM2014}.

The solution of global optimization problems requires several evaluations of objective functions, whose analysis can be time-consuming. Hence, such problems can also benefit from surrogate modeling. In this context, Efficient Global Optimization (EGO) was proposed by \citet{Jones1998}, where a Kriging surrogate is adaptively built in such a way that exploration of design space and exploitation of promising regions are balanced. Updated information regarding the values of objective functions in a-priori selected points and the built-in Kriging variance are considered, when selecting new points to compose the design of experiment of the surrogate model. Thus the so-called Expected Improvement Function is defined (See Appendix A for details). Many recent studies on the subject derive from or directly apply the concept of EGO. \citet{Mohamed2018} combined EGO with the partial least-squared method in order to address higher-dimensional problems.  \citet{Satadru2019} combined EGO with partial least squares and a gradient-based method to develop an optimization algorithm, which was then utilized to solve a complex topology optimization problem. \citet{Guzman2019} addressed design optimization of airframes using EGO in a multidisciplinary optimization context, where critical dynamic aeroelastic loads are estimated and the so-called modal strain energy coefficient is studied as an indicator of the necessity of further exploring the design space, should dynamic aeroelastic loads significantly change. \citet{Ariyarit2018} performed EGO with a multi-fidelity optimization technique applied to design optimization of helicopter blades, searching for maximum blade efficiency. \citet{Carraro2019} proposed an adaptive scheme for selecting target variances, which are then used as parameters to perform EGO in the design of a tuned mass damper.

The consideration of uncertainties further complicates the problem, rendering even higher computational costs. In order to address the complexity of reliability analysis, \citet{Bichon2008} adapts the general procedure of EGO in such a way that limit state equations are replaced by an adaptive Kriging metamodel. Reliability space is efficiently explored, so that the design of experiments is enriched considering both the proximity of candidate points to the limit state equation and the variance of the Kriging surrogate. The resulting technique, known as Efficient Global Reliability Analysis (EGRA) was applied in the solution of RBDO problems \citep{Bichon2009,Bichon2010}. This strategy is thus far utilized in several reliability analysis and optimization studies, either directly or as a key part of novel proposed techniques \citep{LIU2019,chaudhuri2019,Sadoughi2018,Wang2018c}. 

\citet{Wang2018} present an interesting application of adaptive Kriging in time-dependent structural optimization. Despite some similarities with this work, the authors perform RBDO through Stochastic  Equivalent  Transformation \citep{Wang2016}.
	
In the present paper, Risk Optimization is performed in time-dependent problems. The entire trajectories of limit state equations are considered, so that complete information is obtained from the analysis instead of only extreme values. There is no previous work, to the best knowledge of the authors, where adaptive Kriging is used in this context. Thus, a new strategy is proposed herein, where coupled adaptive Kriging surrogate models are iteratively built. The design of experiment of each metamodel is enriched according to different strategies, depending on the nature of model to be surrogated.

\subsection{Proposed Framework }

In this paper, a nested adaptive surrogate modeling approach is proposed, for the first time, for solving time-variant risk optimization problems. The solution scheme employs well-known EGO and EGRA formulations, described in the Appendix, but the application to time-variant risk optimization problems is novel. Moreover, some of the problems addressed herein have not been solved before, as they address stochastic load processes and stochastic strength degradation. The strategy addressed herein involves two nested loops: the inner loop involves determination of cumulative probabilities of failure, which are used as inputs for the objective function, iteratively evaluated as the outer loop searches for the optimum point. The result is a comprehensive Risk Optimization solution framework, suitable to account for time-dependent loads, load-path dependency and structural degradation. In this context, the strategy adopted for solving reliability problems is a key aspect. Considering that most known reliability assessment techniques are imprecise or inadequate in this case, as detailed in Section \ref{sec:Reliab}, a Monte Carlo estimation for cumulative probability of failure is adopted, as given in Equation (\ref{eq:Pfc}). The downside of this approach is increasing the computational cost involved in reliability analysis, which is here just one iterative step of the broader optimization algorithm. Thus, EGRA is adopted in this phase with a tight convergence criterion, enforcing accuracy and alleviating the computational burden. This Kriging surrogate is built in the so-called augmented space, which combines both the design and random variables space, so that one single metamodel can be used to compute the failure probability regardless of the current value of design parameters. This is particularly convenient since reliability analyses must be performed for different design configurations, as another level of surrogate model is assembled in the search for objective function minimum.

\begin{figure}[!ht]
	\centering
	\includegraphics[width=1\linewidth,trim={4cm 3.4cm 10.5cm 3.5cm}, clip]{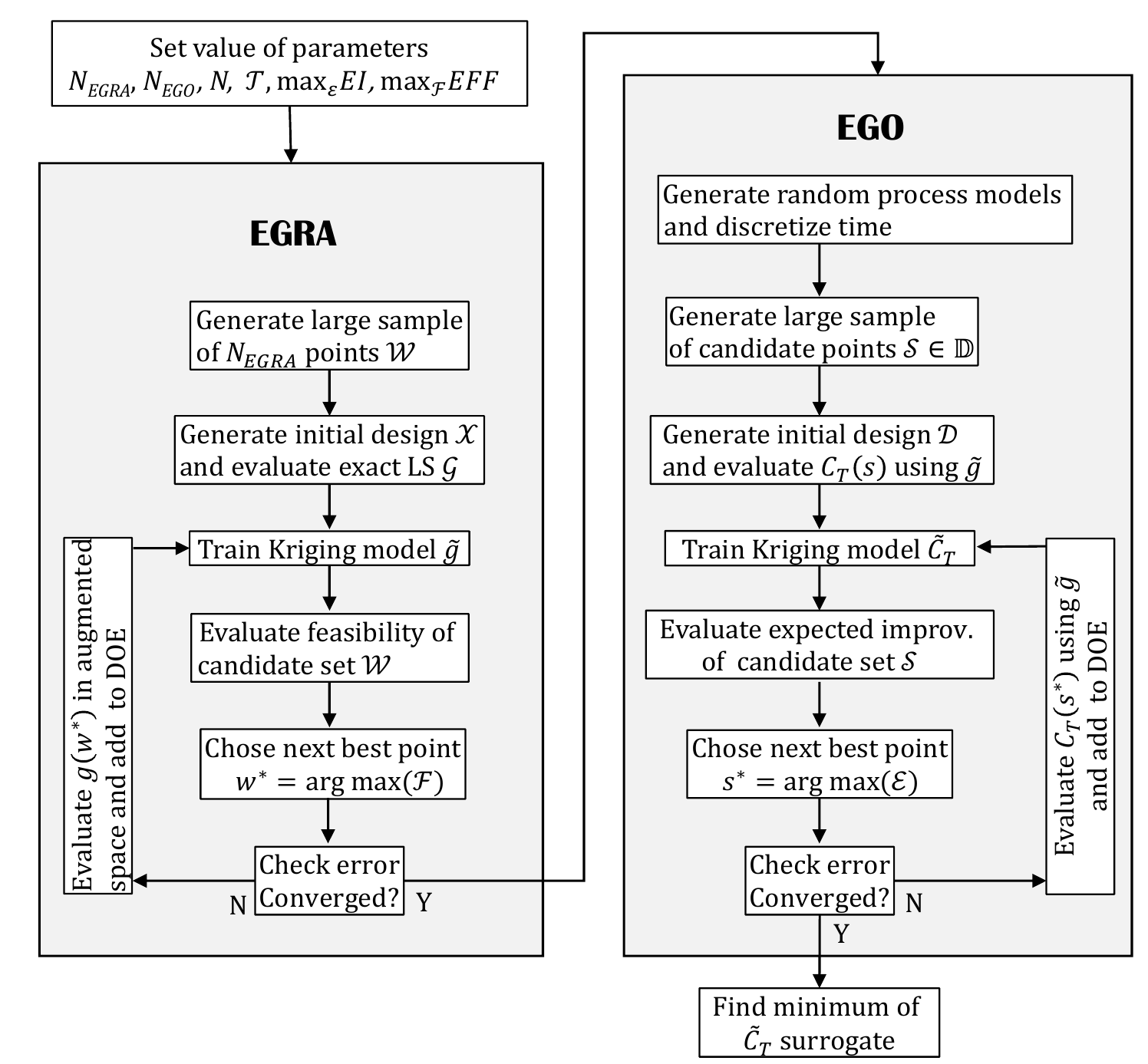}
	\caption{Proposed Framework}
	\label{fig:Scheme}
\end{figure}

The nature of variables which compose the total life-cycle cost in a Risk Optimization problem is very diverse: monetary costs, discount rates, probabilities of failure, material properties, and so on. Hence, the behavior of the objective function tends to be problem-specific. A general technique to address this type of problems must be able to efficiently scan the design space, especially because each design point evaluation defines new limit state configurations. Thus, EGO is adopted to search for the global minimum. This way, an efficient multi-level application of adaptive surrogate models is proposed to address the complexity of time-variant risk optimization problems. The proposed framework is illustrated in Figure \ref{fig:Scheme}

As shown in Figure \ref{fig:Scheme}, the proposed scheme is a combination of EGRA and EGO, with a Monte Carlo scheme for solving the time-variant reliability problem. Hence, the formulations in Sections \ref{sec:TVRPS} and \ref{sec:RO} are an inherent part of the solution scheme proposed in this paper. 

An important feature of the proposed framework is that the complete time series of involved stochastic processes is obtained by simulation, with the only simplification being time discretization. This allows problems like load-path dependency to be addressed, as illustrated in Example 3. In the next section, three novel examples of risk optimization considering random load processes and random strength degradation processes are addressed. To the best of the authors knowledge, no similar  examples have been solved before in a context of risk optimization. Examples 1 and 2 are based on the literature, but the original references only address time-invariant \citep{BlatmanPEM2010} or time-variant \citep{Sudret2008d} reliability analysis.


\section{Examples}
\subsection{Steel beam subject to corrosion}
Consider a steel bending beam with rectangular cross-section $ \acc{b_0,\, h_0}^T $ and length $ L=5m$, submitted to dead loads $  \rho_{st}b_0h_0$ (Nm$^{-1}$), where $  \rho_{st}=78.5$ kNm  is the steel mass density, as well as a pinpoint load F applied at midspan.
\begin{figure}[!ht]
	\centering
		\includegraphics[width=1\linewidth,trim={4cm 9cm 5cm 7cm}, clip]{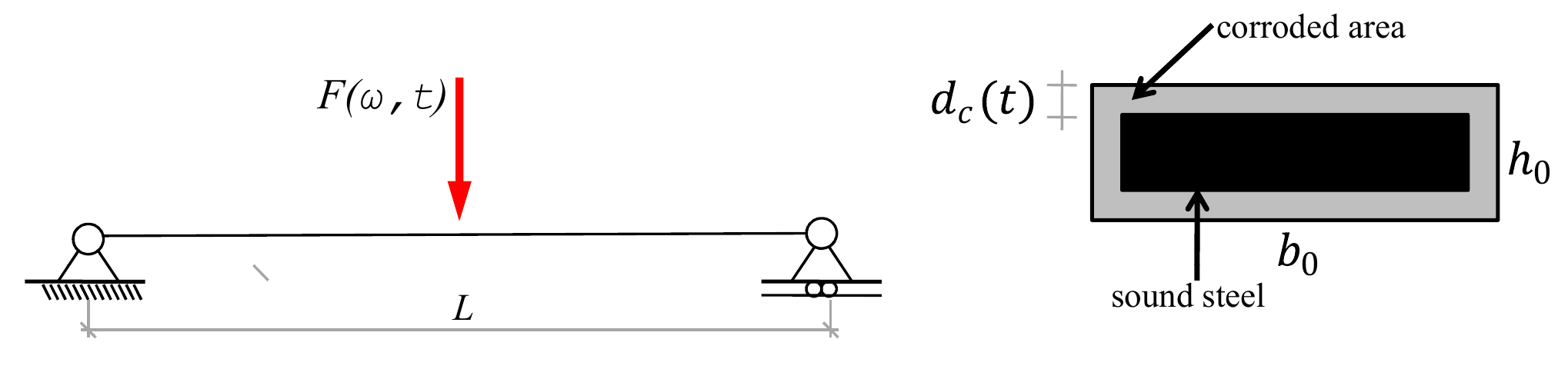}
	\caption{Corroded beam under a midspan load, after \citep{Sudret2008d}}
	\label{fig:SteelBeam}
\end{figure}

\begin{figure}[!ht]
	\centering
	\subfloat[Fixed
        $\kappa$]{\label{fig:beamConvergence1}\includegraphics[width=0.3\textwidth]{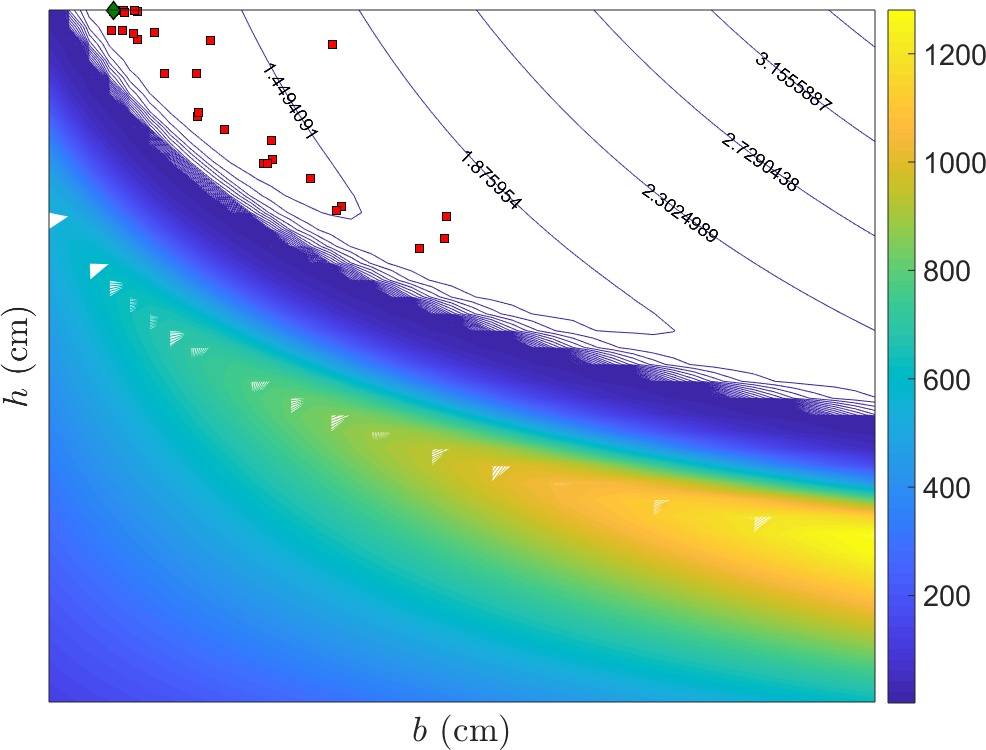}}%
	\hfill
	\subfloat[$\kappa$ as a random
        variable]{\label{fig:beamConvergence2}\includegraphics[width=0.3\textwidth]{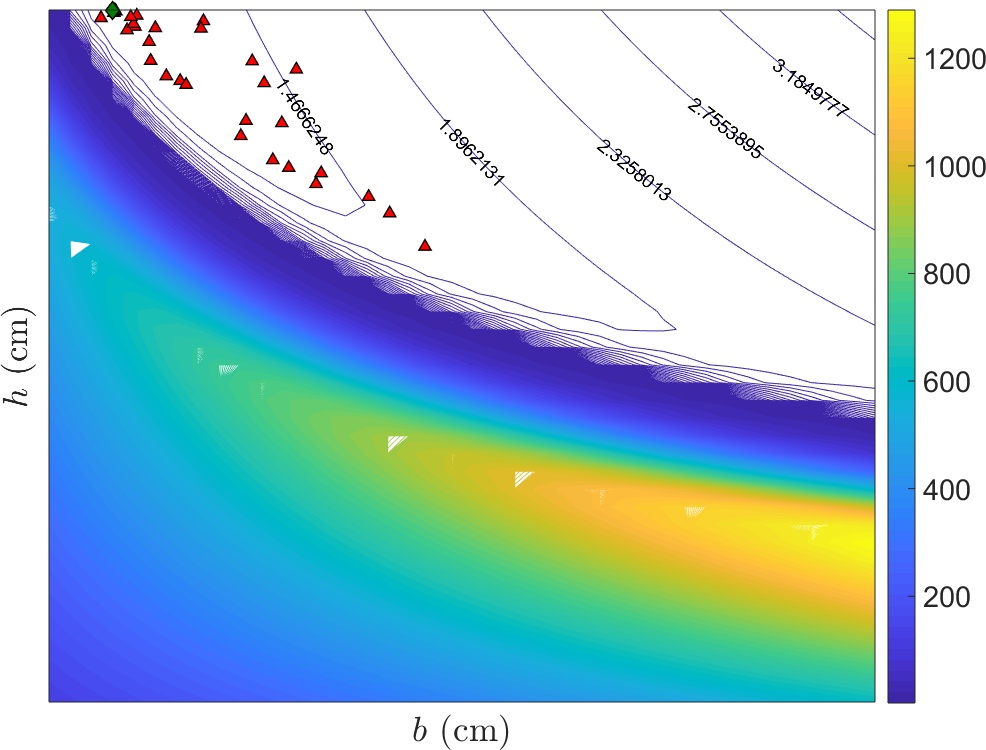}}%
	\hfill
	\subfloat[$\kappa$ as a stochastic
        process]{\label{fig:beamConvergence3}\includegraphics[width=0.3\textwidth]{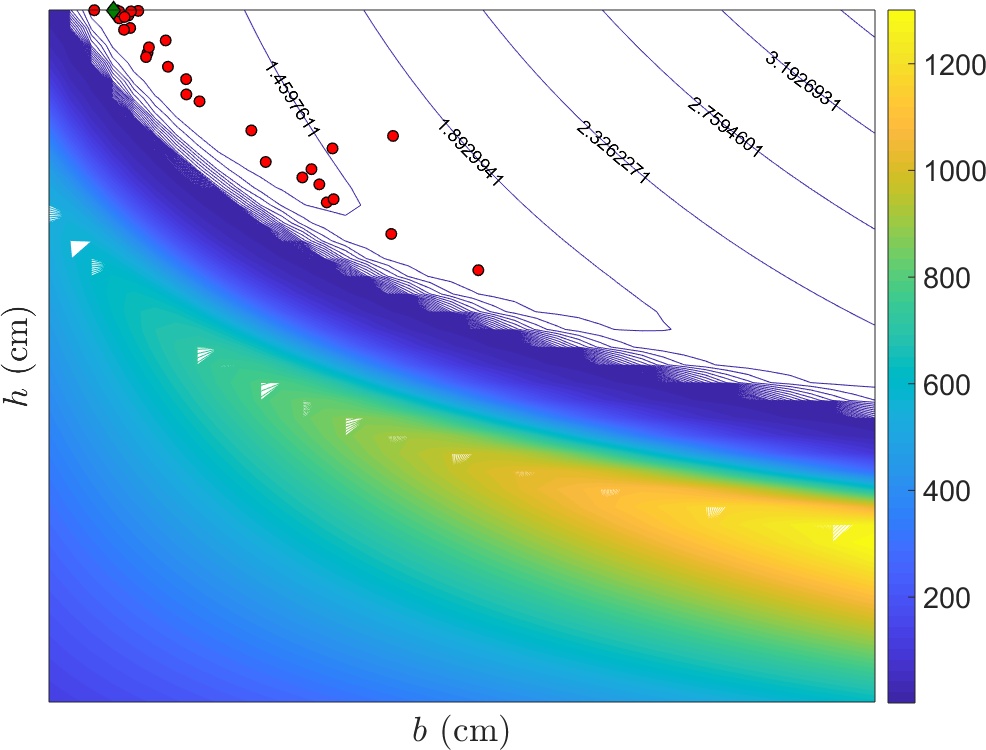}}%
	\caption{Contour plot of the total costs for different design configurations. Red marks represent the solutions for $30$ replications considering different cases for the beam example, while green marks represent the PSO solutions for each case.}
	\label{fig:beamConvergence}
\end{figure}

The beam is also subjected to corrosion, in such a way that the corrosion depth $ d_c $ in all the faces of the beam increases linearly with time, \ie{} $ d_c=\kappa t $. Moreover, it is assumed that the corroded areas have lost all mechanical stiffness. The limit state function which describes the formation of a plastic hinge at midspan reads: 
\begin{equation} \label{eq:corrosion}
g(\ve{d},t,\textbf{X})= \dfrac{(b_0 - 2\kappa t)(h_0 - 2\kappa t)^2 f_y}{4} -(\dfrac{FL}{4}+\dfrac{\rho_{st}b_0h_0L^2}{8}),
\end{equation}
where the yield stress is denoted by $f_y$. The analysis is carried out considering the time interval [0, 10] years. Three corrosion scenarios are considered. In the first one, corrosion kinetics is controlled by deterministic $ \kappa=1 $  mm year$^{-1}$. The second scenario considers the corrosion rate as a random variable with mean of $1$ mm year$^{-1}$ and a coefficient of variation of 30\%. In the third scenario, the corrosion rate is considered as a discrete pulse process, with annual renewal, and mean intensity of $ 1$ mm year$^{-1}$ and coefficient of variation of 30\%. In all scenarios, the load $F$ is modeled as a Gaussian random process with mean $ 6 \ kN $, coefficient of variation $ 0.3 $ and a Gaussian autocorrelation function with correlation length $ \lambda =1 $ month. For all scenarios, the same inner surrogate model $\widetilde{g}$ is used, \ie a part of the computational cost for obtaining a solution for different corrosion scenarios can be reduced. The random parameters are gathered in Table~\ref{tablebeam}. The risk optimization problem is defined by Eq.~\eqref{eq:beamRO}

\begin{equation}
\begin{aligned}
&C_T = C_I + \sum_{i=1}^{10} C_f P_{{fc}_i} \\ 
&\text{s.t. } 0.1 \leq b_0 \leq 0.5  \\
&\text{\,\,\,\,\,\,\,\,\,\,} 0.01 \leq h_0 \leq 0.06
\end{aligned} 
\label{eq:beamRO}
\end{equation}
The initial costs are related to the cross section of the beam $C_I = \nu b_0 h_0$, with $\nu=1/125$, and the failure costs are considered to be $1,000$ times higher, \ie{} $C_f = 1,000 \, C_I$. In this academic example, a very high cost of failure is adopted, penalizing the unsafe regions of the design space, generating a total cost function with two very distinct regions (a high $ C_T $ values region and low $ C_T $ near-plateau one), separated by a very steep transition, as shown in Figures \ref{fig:beamConvergence1}, \ref{fig:beamConvergence2} and \ref{fig:beamConvergence3}. Approximating this objective function is a significant challenge to the Kriging approximation.  A monthly discount rate of $1\%$ is also considered. The optimization problem consists in determining $\ve{d} = \acc{b_0, \,h_0}^T$ that minimizes the total cost $ C_T(\ve{d})$.
\begin{table}[!ht]
	\centering
	\caption{Corroded beam random variables and parameters}
	\label{tablebeam}
	\begin{tabular}{llll}
		\hline
		Parameter          & Distribution & Mean   & COV  \\ \hline
		Steel yield stress (MPa) & Lognormal    & 240 & 10\% \\
		Beam breadth (m)       & Lognormal    & $ b_0 $   & 3\%  \\
		Beam height (m)       & Lognormal    & $ h_0 $  & 3\% \\
		\hline
	\end{tabular}
\end{table}

\begin{figure}[!ht]
	\centering
	\includegraphics[width=0.5\linewidth]{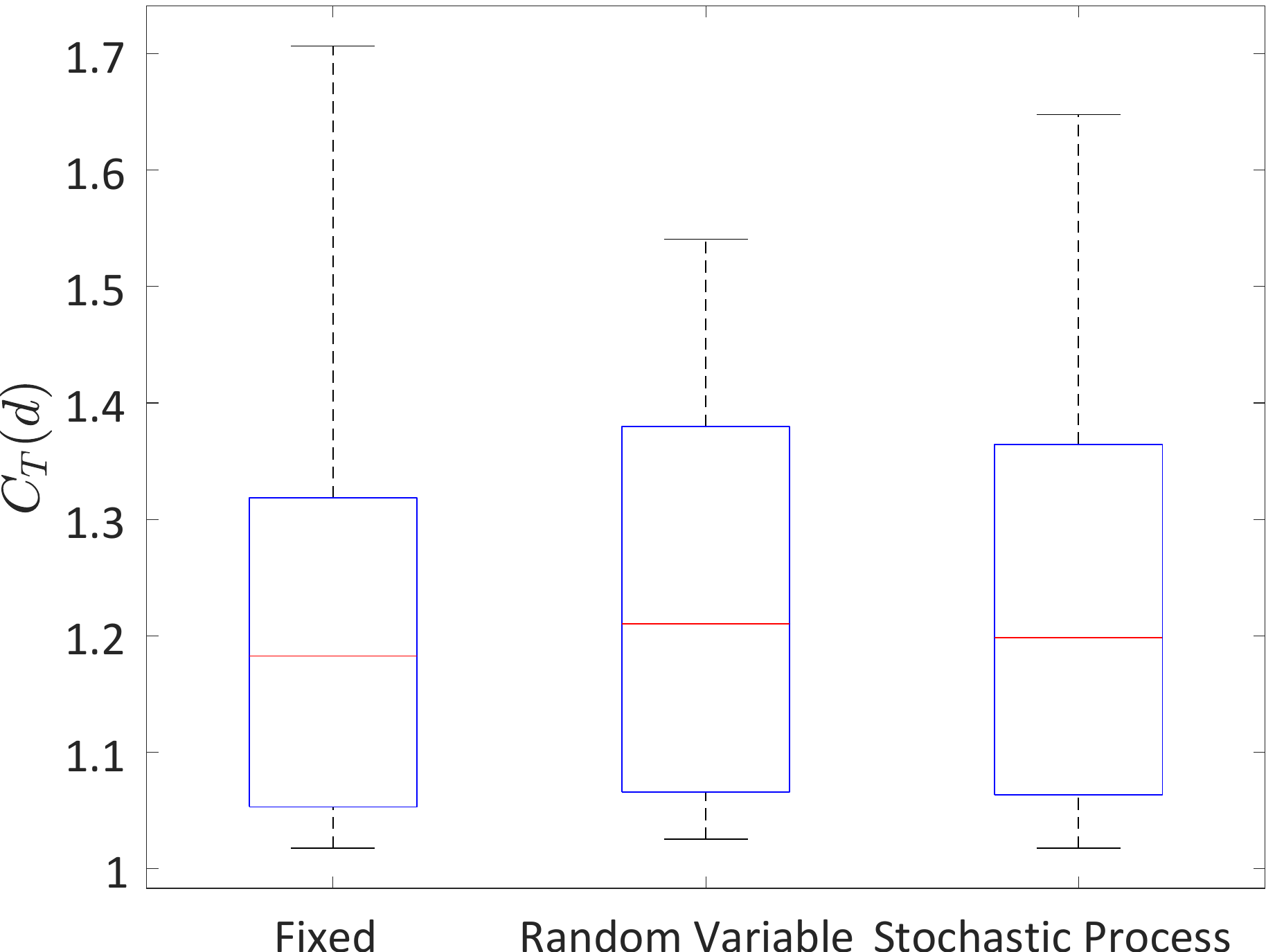}
	\caption{Box-plot for total costs considering different modeling of the corrosion rate $\kappa$. (30 replications of the analysis)}
	\label{fig:CostBoxPlotBeam}
\end{figure}

Figures \ref{fig:beamConvergence1}, \ref{fig:beamConvergence2}, and \ref{fig:beamConvergence3}  show contour plots of the cost functions for the three corrosion scenarios. Note that the plane regions have very few contours, while in the steep regions the concentration of contours is very high. The red squares, triangles and circles are the results of 30 optimization runs for each case. A Particle Swarm Optimization (PSO) \citep{Bansal} is also performed on the problem without the aid of surrogate models, considering 30 particles per iteration in order to compare the results. The stopping criteria for the PSO is a tolerance of $ 10^{-4} $ in the change of the value of all design variables. This result is represented by the green diamond and serves here as reference. On average, for fixed, random variable and stochastic process corrosion rate, the objective function was called 23, 25 and 27 times, respectively, in the solutions using the EGO approach, and 480, 510 and 510 times in the solutions using the PSO approach. The results for total costs are compared in Figure \ref{fig:CostBoxPlotBeam}. The box-plots show the optimal results for $ C_T $  obtained in the 30 runs of each case. Fairly precise results can be obtained using the proposed methodology, with a much smaller number of objective function evaluations. The stochastic corrosion process version of this problem had never been solved before.

%
%
%

\subsection{23-bar Plane Truss}
\begin{figure}[!ht]
	\centering
	\includegraphics[width=0.85\linewidth,trim={4.1cm 8cm 4.5cm 6.2cm}, clip]{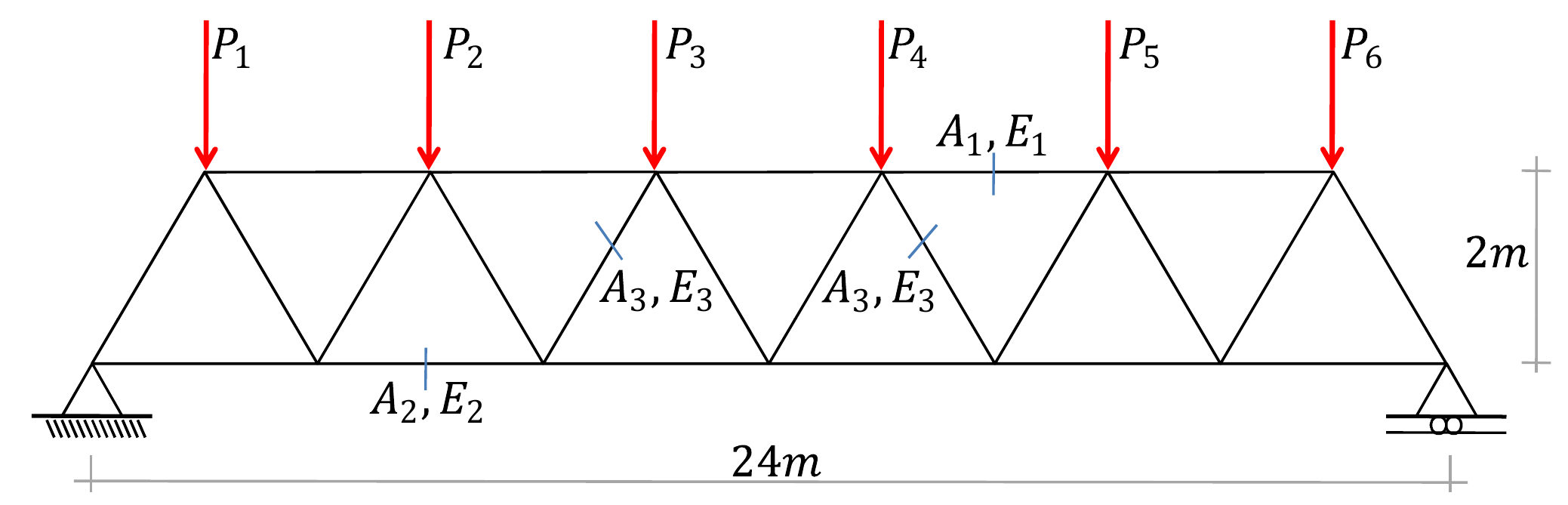}
	\caption{Corroded beam under a midspan load, adapted from \citep{BlatmanPEM2010}}
	\label{fig:Truss}
\end{figure}

\begin{figure}[!ht]
	\centering
	\includegraphics[width=0.5\linewidth,trim={7cm 7.1cm 9.4cm 5.85cm}, clip]{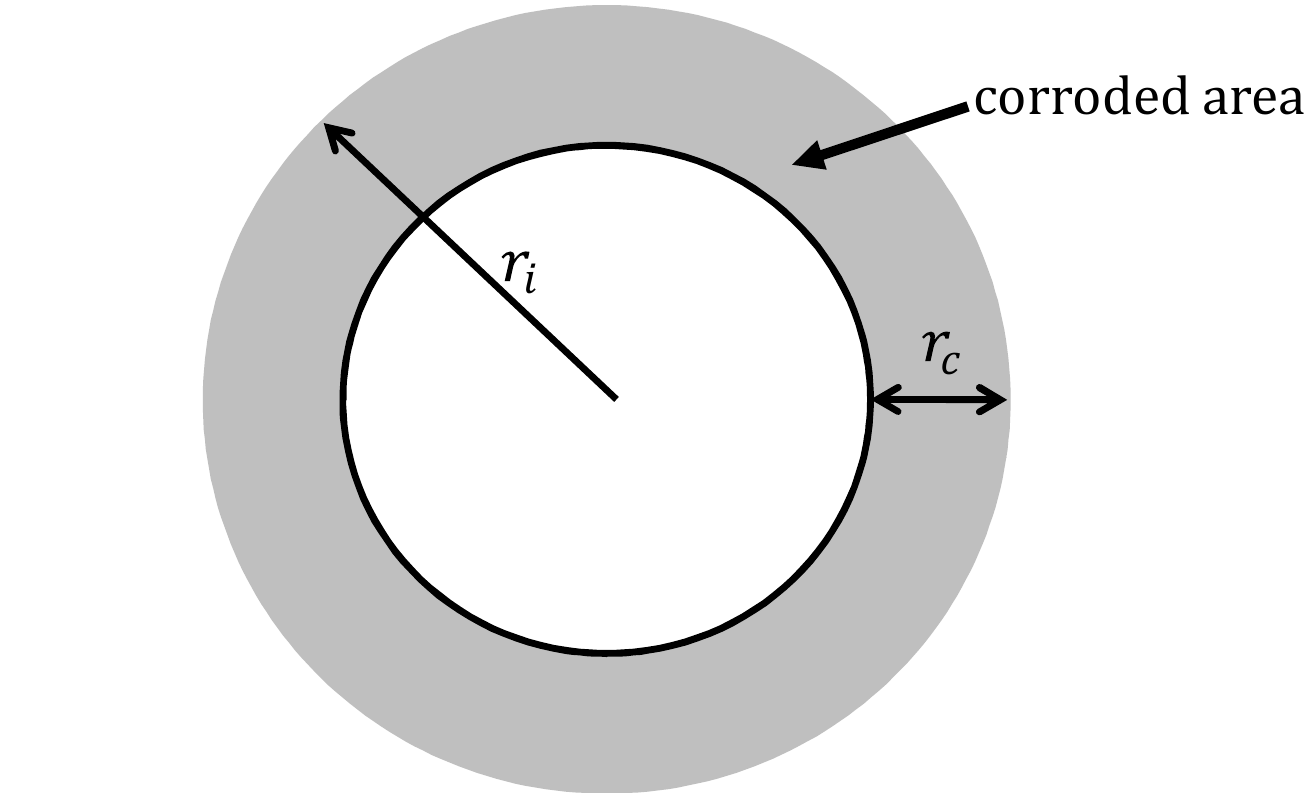}
	\caption{Cross section} 
	\label{fig:CrossSectionTruss}
\end{figure}

A 2D truss structure is considered, as shown in Figure \ref{fig:Truss}. It is composed by 23 bars and 13 nodes, and subjected to six vertical time-varying loads applied on the upper nodes. The magnitudes of all vertical loads are cast as a single stationary Gaussian process with mean value $50$ kN, standard deviation $7.5$ kN and Gaussian autocorrelation coefficient function with a correlation length of $ \lambda = 1 $ year.  There are three types of bars, with different cross-sectional areas and materials, as indicated in Figure \ref{fig:Truss}. Circular bars are considered, and the design variables  $d_1, d_2$ and $d_3$ are the radius of the three types of bars.
The bars are subjected to corrosion, so that the radius of the cross section is decreased over time, following $ r_c=\kappa t $. The radius at $t=0$ is $r_i$, and the current radius at any given time is $r(t) = r_i - r_c(t) $ as shown in Figure \ref{fig:CrossSectionTruss}.

\begin{figure}[!ht]
	\centering
	\subfloat[Case $\#1$]{\label{fig:BoxPlotCase1}\includegraphics[width=0.33\textwidth]{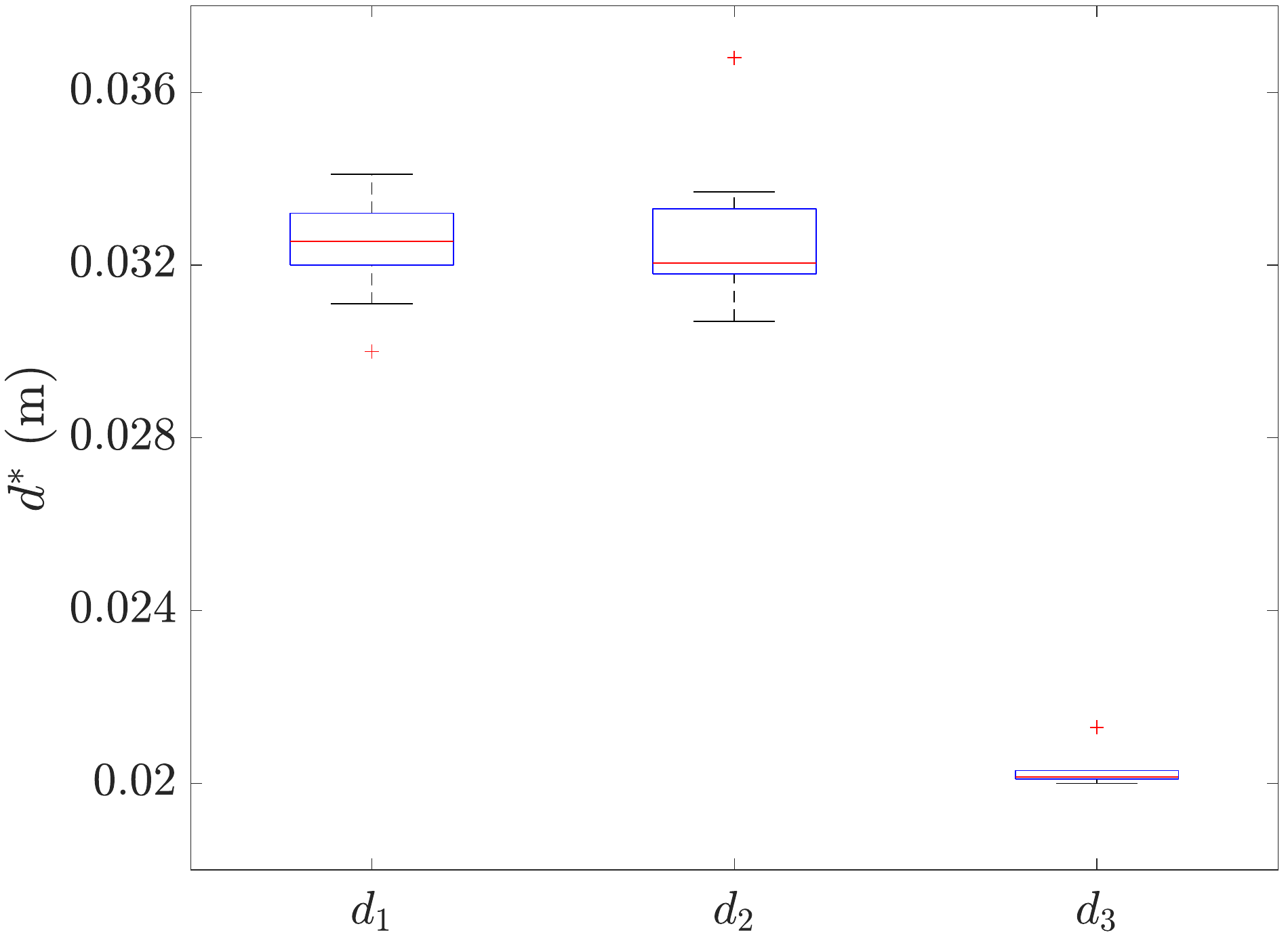}}%
	\hfill
	\subfloat[Case $\#2$]{\label{fig:BoxPlotCase2}\includegraphics[width=0.33\textwidth]{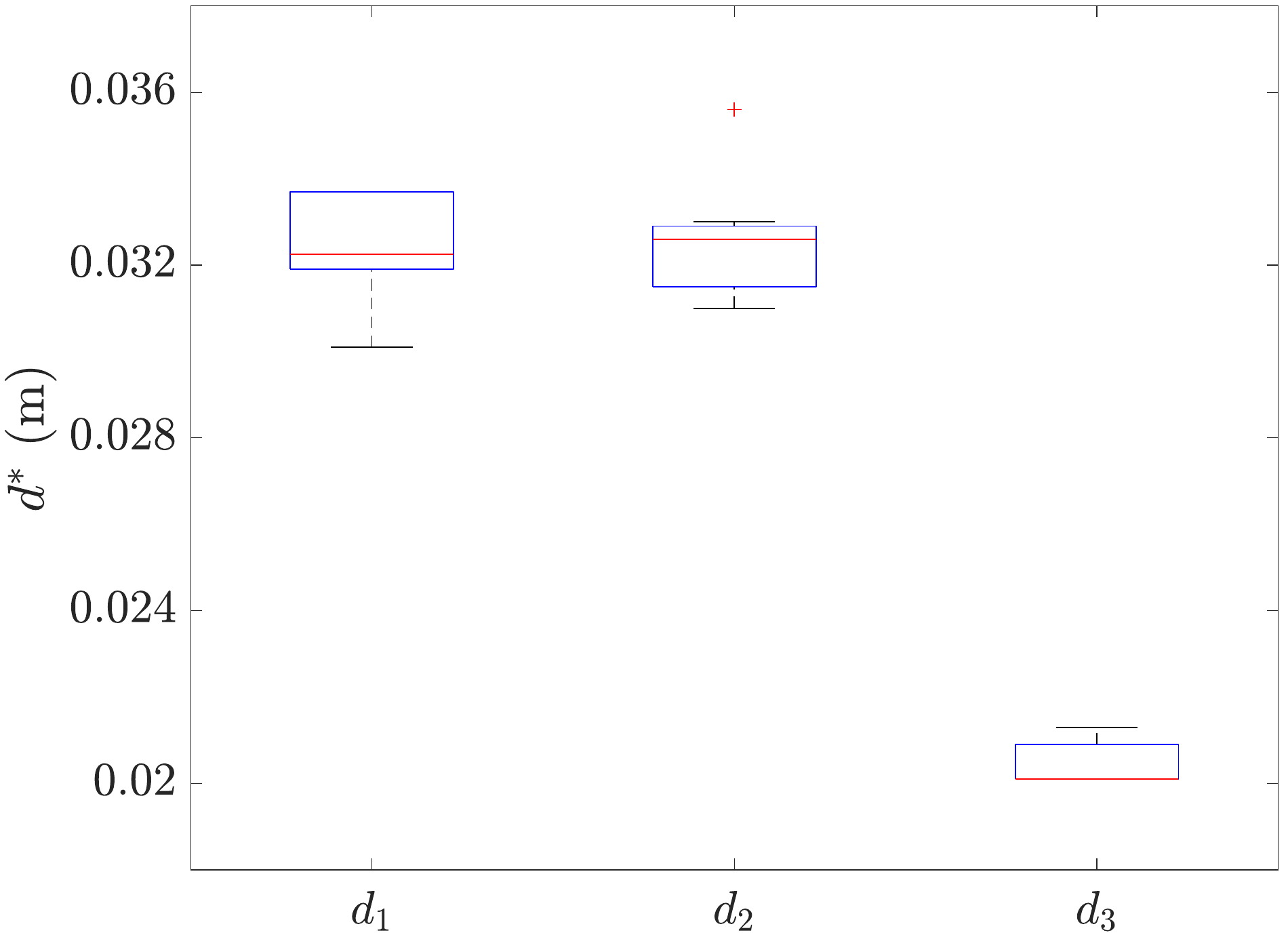}}%
	\hfill
	\subfloat[Case $\#3$]{\label{fig:BoxPlotCase3}\includegraphics[width=0.33\textwidth]{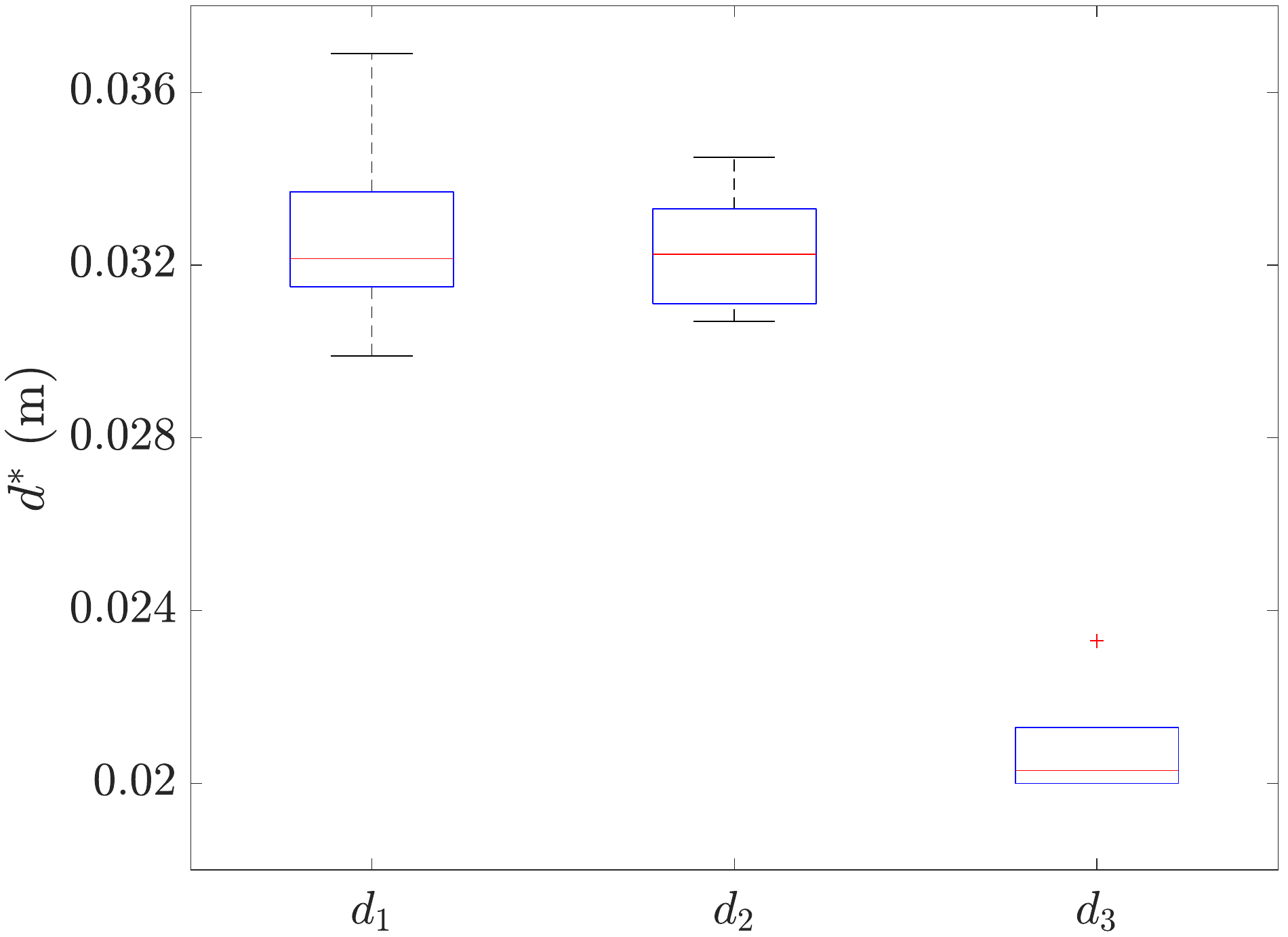}}%
	\caption{Design solutions with $20$ replications considering different cases for the truss example.}
	\label{fig:optim_d}
\end{figure}

\begin{figure}[!ht]
	\centering
	\includegraphics[width=0.5\linewidth]{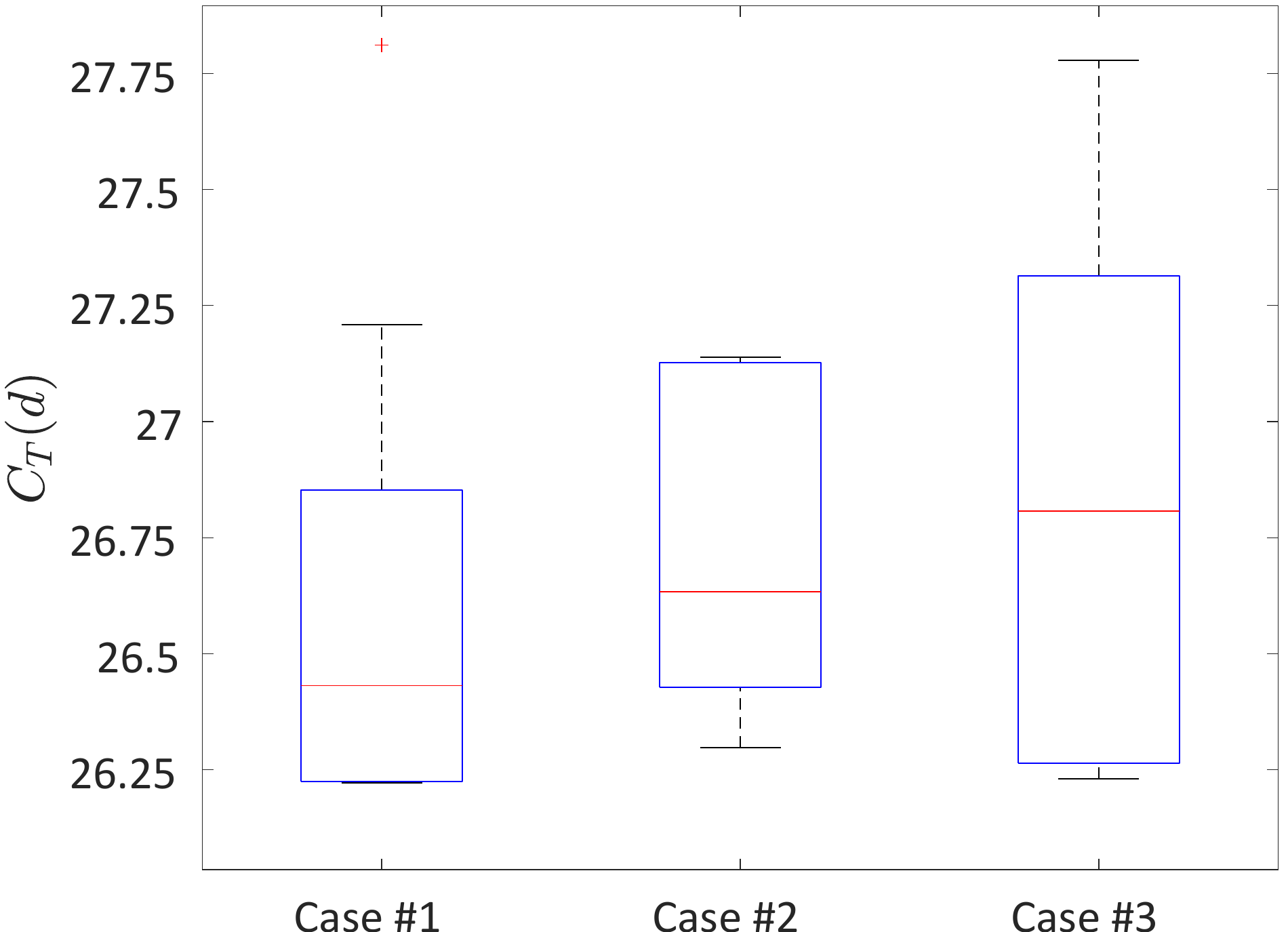}
	\caption{Optimum costs obtained for $20$ runs of each corrosion rate case.}
	\label{fig:CostBoxPlot}
\end{figure}

 Three scenarios are considered for corrosion rate: in case $\#1$, a deterministic $ \kappa = 10 \, \mu m \, year^{-1} $ is considered. In case $\#2$, three different corrosion rates are defined, one for each type of bar. $ \kappa_1^{RV}, \kappa_2^{RV} $and $\kappa_3^{RV} $ are correlated random variables with mean 10 $\, \mu m \, year^{-1} $ and COV of 30\%. The correlation coefficient is set to 0.8. In case $\#3$, three discrete pulse processes are used to model the three corrosion rates of the different types of bars. The processes have annual renewal, with mean $10 \mu m $ and COV of 30\% and correlation coefficient between them of 0.8. Table~\ref{tabletruss} describes the remaining random variables of the problem. The limit state equation is defined implicitly by a finite element model, and is written in terms of the vertical displacement of the mid-span node, herein denoted by $V_1$. A maximum allowed displacement of $ 0.1 $ m is considered:
\begin{equation} \label{eq:LStruss}
g(\ve{d},t,\textbf{X})= 0.1 - V_1(\ve{d},t,\textbf{X}).
\end{equation}
The time interval in which the analysis is carried out is [0, 30] years, so that the formulation of the risk optimization problem reads:

\begin{equation}
\begin{aligned}
&C_T = C_I + \sum_{i=1}^{30} C_f P_{{fc}_i}, \\ 
&\text{s.t. } 0.02\, \text{m} \leq d_1 \leq 0.04\, \text{m}  \\
&\text{\,\,\,\,\,\,\,\,\,\,} 0.02\, \text{m} \leq d_2 \leq 0.04\, \text{m} \\
&\text{\,\,\,\,\,\,\,\,\,\,} 0.02\, \text{m} \leq d_3 \leq 0.04\, \text{m}
\end{aligned} 
\label{eq:trussRO}
\end{equation}
The design costs are proportional to the area of the bars, \ie $C_I = 10^4(d_1^2 + d_2^2 + d_3^2) $, and the cost of failure is obtained as $ C_f = 10 C_I $.  An annual discount rate of $1\%$ is also considered. The results of optimum cost for 20 analyses are summarized in Figures \ref{fig:optim_d} and \ref{fig:CostBoxPlot}. Figure \ref{fig:optim_d} shows the boxplot for each design variable on each case. Figure \ref{fig:CostBoxPlot} gathers the corresponding optimum total costs. The methodology provides consistent solutions. On average, only 13, 13 and 14 cost function evaluations were needed to reach a solution for each corrosion rate scenario, respectively.

\begin{table}[!ht]
	\centering
	\caption{ Corroding truss random variables and parameters}
	\label{tabletruss}
	\begin{tabular}{llll}
		\hline
		Parameter          & Distribution & Mean   & COV  \\ \hline
		$ E_1 (MPa)  $            & Lognormal    & 210,000   & 10\% \\
		$ E_2 (MPa)  $            & Lognormal    & 210,000   & 10\%  \\
		$ E_3 (MPa)  $            & Lognormal    & 210,000   & 10\%  \\
		$ A_1 (cm^2) $            & Lognormal    & $ \pi r_1^2 $       & 10\% \\
		$ A_2 (cm^2) $            & Lognormal    & $ \pi r_2^2 $       & 10\% \\
		$ A_3 (cm^2) $            & Lognormal    & $ \pi r_3^2 $       & 10\% \\
		\hline
	\end{tabular}
\end{table}

\FloatBarrier
\subsection{Load-path dependent Truss}

\begin{figure}[H]
	\centering
	\includegraphics[width=0.9\linewidth,trim={0cm 5.5cm 0cm 6cm}, clip]{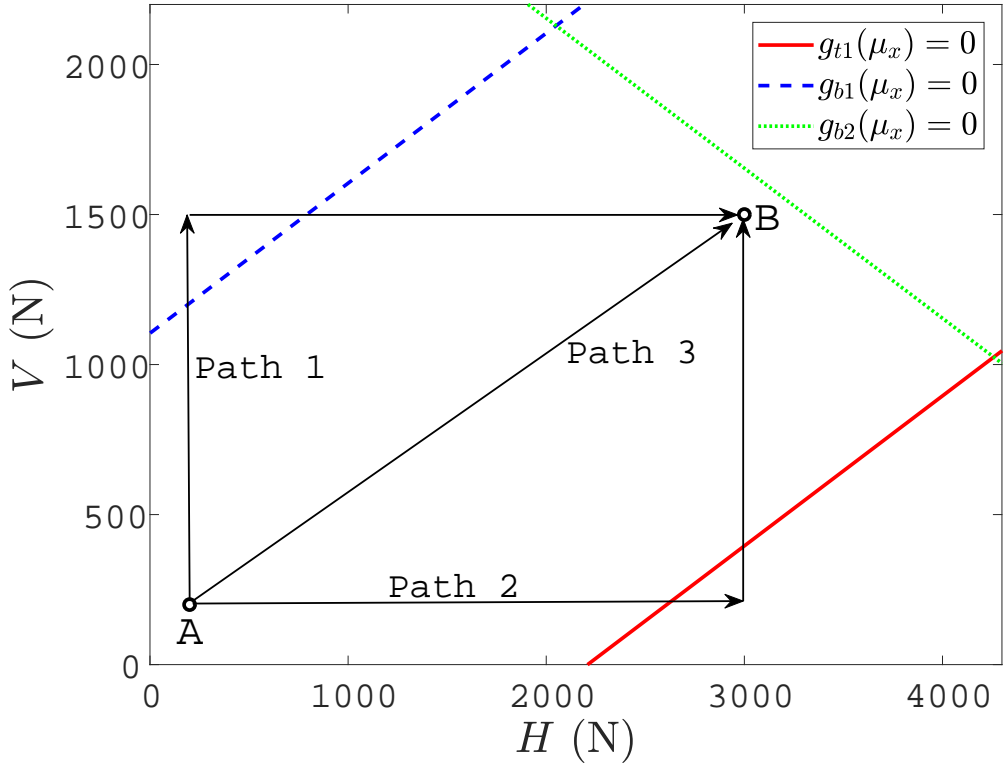}
	\caption{Load Paths}
	\label{fig:AllPaths}
\end{figure}

Consider the truss composed by circular bars 1 and 2, as shown in Figure \ref{fig:2BT}. Two time-variant loads $ H(t) $ and $ V(t) $ are applied on the upper node.  Three failure modes are considered: tensile rupture of bar $1$ $(g_{t1})$, buckling of bar $1$ $(g_{b1})$, and buckling of bar $2$ $(g_{b2})$. Thus, a time-variant system reliability problem is defined considering the limit state equations associated to these failure modes:

\begin{figure}[h]
	\centering
	\includegraphics[width=0.7\linewidth,trim={7.7cm 10.2cm 10.7cm 3.3cm}, clip]{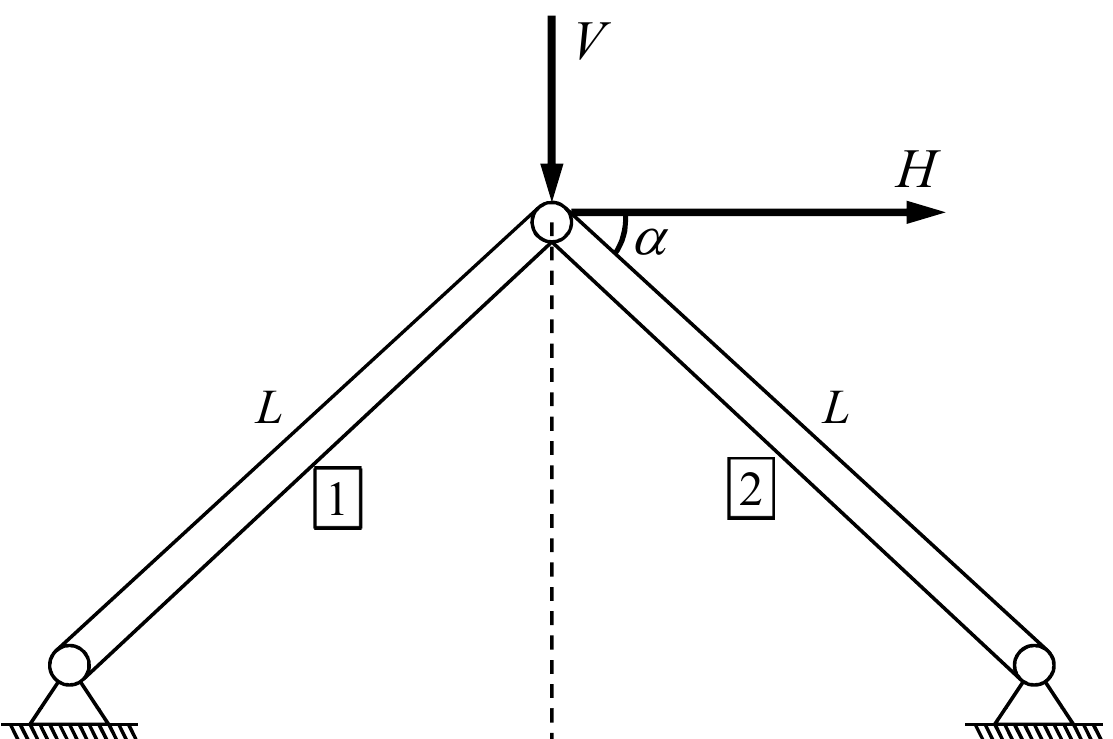}
	\caption{Two-bar truss scheme}
	\label{fig:2BT}
\end{figure}

\begin{equation*} \label{eq:t1}
\begin{aligned}
& g_{t1}(\bm{X},t) = A_1 \sigma_u - \Bigg[  \dfrac{H(t)}{2\cos \alpha} - \dfrac{V(t)}{2\sin \alpha} \Bigg] \\
& g_{b1}(\bm{X},t) =   \dfrac{\pi^2EI_1}{L^2}- \Bigg[ - \dfrac{H(t)}{2\cos \alpha} + \dfrac{V(t)}{2\sin \alpha} \Bigg]    \\
& g_{b2}(\bm{X},t) =  \dfrac{\pi^2EI_2}{L^2}-\Bigg[ \dfrac{H(t)}{2\cos \alpha} + \dfrac{V(t)}{2\sin \alpha} \Bigg]    \\
& g_{sys}(\bm{X},t) = \min (g_{t1},g_{b1},g_{b2})
\end{aligned}
\end{equation*}
where $A_i$ is the area of the $i$-th bar in $m^2$ and $L$ is the length of the bars in $m$. The truss is symmetric. The two bars have the same Young Modulus $E$, defined as a normal random variable with $\mu_E=70$GPa and ${COV}_E=0.03$, and the same ultimate tensile strength, defined as a normal random variable $\sigma_u$, with $\mu_{\sigma_u} =24.5643$MPa and ${COV}_{\sigma_u}=0.1$. This value of ultimate stress was set so as to result in a tight compromise between the three different failure modes. The probability that random variables reach negative values is very small and can be neglected in this example. This problem is load-path dependent, i.e. the structure can violate different limit states or fail at different times depending on the trajectory that the loads follow in time. To illustrate the load path dependent problem, consider that the radius of the cross sections are $r_1 = 4 $mm for the first bar, and $r_2 = 5.2 $mm for the second bar. Figure \ref{fig:AllPaths} shows three possible load paths, as well as the limit state equations, evaluated at the mean $ \mu_{\bm{X}}$. Suppose that at time $t=t_0$ the loads are at point $A$, and at $t=t_f>t_0$, the loads correspond to point $B$. If the loads follow Path $1$, the structure fails due to buckling of the first bar. If the loads follow Path $2$, the horizontal load is increased first, and the structure fails by tensile rupture of bar 1. Now, if the loads follow Path 3, which corresponds to a concomitant increase in both loads, point $B$ is safely reached, and there is no failure. Thus, the load-path dependency of the problem is demonstrated.

When the loads are stochastic processes, there is an infinite number of possible trajectories, and evaluating structural reliability depends on considering such trajectories, which only adds complexity to the problem. Load-path dependent problems cannot be solved by usual techniques, such as time-integration (extreme value analysis) or load combination, as discussed in \citet{Melchers2018}. However, load path-dependent reliability problems can be solved by explicit simulation of load process realizations, as proposed in this work. 

Consider now that one is interested in the optimal areas for the two bars, aiming at minimizing total costs in a risk-optimization scenario. Forces $V(t)$ and $H(t)$ are stochastic Gaussian processes with means $1$~kN and $2$~kN, respectively. Both loads have a COV of 0.2 and a correlation length of $\lambda_V = \lambda_H = 1$ month. The auto-correlation function of the random processes is given by:

\begin{equation}
R(x,\lambda)=\exp\bigg[-\Big(\frac{x}{\lambda}\Big)^2\bigg]
\end{equation}

The loads are independent of each other and of the other random variables. A time interval of 10~years is studied, so that the objective function of the problem can be stated as:

\begin{equation}
\begin{aligned}
&C_T(r_1,r_2) = C_I(r_1,r_2) + \sum_{i=1}^{10} C_f P_{{fc}_i}(r_1,r_2) \\ 
&\text{s.t. } 4 \, \text{mm}  \leq r_1 \leq 6 \, \text{mm}  \\
&\text{\,\,\,\,\,\,\,\,\,\,} 4 \, \text{mm} \leq r_2 \leq 6 \, \text{mm}
\end{aligned} 
\label{eq:LPDRO}
\end{equation}

The initial costs are proportional to the volume of the structure $C_I(r_1,r_2)=10^5\big(A_1(r_1)+A_2(r_2)\big)L$, and the cost of system failure is 10 times higher. An annual discount rate of $2\%$ is also considered. Different failure costs could be associated to different  limit states, without any change in the solution procedure. 

Table 3 shows the results for the optimization problem, comparing 10 runs of the approach proposed in this work (denoted by 'EGO') and a reference obtained with 20 generations of 30 particles of a PSO algorithm, performed without the aid of surrogate models. The standard deviations of the obtained results are denoted between parenthesis.

\begin{table}[!ht]
	\centering
	\caption{ Mean and COV of optimization results and reference}
	\label{tableLPD}
	\begin{tabular}{lllll}
		\hline
		& $r_1$(mm) & $r_2$(mm)   & $C_T$ & $N_{calls}$ \\  	\hline
		EGO      & 4.37(0.01)    & 5.32(0.01)  & 5.16(0.01) & 17(5.1) \\
		PSO      & 4.35    & 5.29  & 5.13 & 600  \\
		
		\hline
	\end{tabular}
\end{table}

As seen from table 3, the results obtained with both methodologies are remarkably consistent, with less than $1\%$ discrepancy between the optimum design radii and associated total cost.

\section{Conclusion}
Expected life-cycle cost, or risk optimization, allows one to find the optimal points of compromise between safety and economy in structural desing. Typically, the underlying reliability problem is time-variant, and its solution is far from trivial. Problems involving strength degradation or load-path dependency usually require solution by Monte Carlo simulation, with a large computational burden, especially in an optimization context. To address efficiently and accurately this type of problem, a nested Kriging approach with active learning is proposed in this paper. The strategy is based on constructing two adaptive Kriging surrogates. One surrogate is built so as to mimic the objective (cost) function, starting from a design of experiment built with LHS in the space of the design variables, which is further enriched as the optimization problem is solved using the EGO approach. Another Kriging surrogate model is built for each limit state function, starting from a first design of experiment built with LHS in the augmented space of both design and random variables. The surrogate is then enriched using the EGRA strategy.
 
Three novel risk optimization problems have been addressed, involving stochastic process strength degradation and stochastic process loading. These problems considered analytical and numerical (finite element) limit states. A complex load-path dependent problem was also addressed for the first time in an optimization context. Satisfactory accuracy and convergence was observed in all examples, with a few calls to the objective function. Solution cost was shown to be approximately the same for three different models of degradation involving a deterministic corrosion rate, a rate modeled by random variables, and by a random process.

On the other hand, the number of evaluations of the inner surrogate model was found to be excessively large for this strategy to be applied in problems that combine extremely low failure probabilities together with time series that require a large number of discretization points. Further studies are necessary in order to adapt the method to this kind of problems, and to increase the scope of the solution to involve dynamic problems that cannot be represented by pointwise surrogates of the limit state equations. The proposed technique was capable of solving problems involving stochastic strength degradation, time-dependent loads and load-path dependency.

\section{Acknowledgment}
The first author thanks the Brazilian Council for Higher Education (CAPES) for funding a six-month research term at ETH Zurich "Programa de Doutorado Sandu\'{i}che no Exterior", through grant number: 88881.133186/2016-01. The third author aknowledges funding by CNPq (grant n. 306373/2016-5) and FAPESP (grant n. 2017/01243-5).

\nocite{Bourinet2011}

\bibliographystyle{chicago}
\bibliography{usedbib} 

 \appendix
 \section{Kriging Basics}
Kriging \citep{Santner2003}, also known as Gaussian process regression in machine learning \citep{Rasmussen2006}, is an emulator that considers the computational model to approximate as one realization of an underlying Gaussian process:
\begin{equation} \label{eq:KrigDef}
\mathcal{M}(\ve{x})  = \sum_{j=1}^{p} \beta_j f_j(\ve{x}) + Z(\ve{x}),
\end{equation}
where the two summands are a deterministic mean known as the \emph{trend} and a zero-mean stationary Gaussian process, respectively. The trend can take multiple forms yielding different types of Kriging. Universal Kriging corresponds to the most general case when the trend is cast as a linear combination of a collection of $p$ weights $\ve{\beta} = \acc{\beta_j, j = 1 \enum p}$ and regression functions $\ve{f} = \acc{f_j, j = 1 \enum p}$. Ordinary Kriging, which corresponds to the special case when $  p = 1 $ and $f(\ve{x}) = 1$, is considered here. The Gaussian process $Z\prt{\ve{x}}$ is defined by its auto-covariance function $\text{Cov}\bra{Z\prt{\ve{x}}, Z\prt{\ve{x}'}} = \sigma^2 R\prt{\ve{x},\ve{x}';\ve{\theta}}$ where $\sigma^2$ is the Gaussian process variance and $R$ is an auto-correlation function with hyperparameters $\ve{\theta}$. The auto-correlation function is chosen based on some assumptions about the degree of smoothness and regularity of the underlying model. In this work, the Mat\'ern $5/2$ auto-correlation function is considered. It can be defined in the one-dimensional case as:
\begin{equation}
R\prt{x, x^\prime; \theta} = \prt{1 + \sqrt{5} \frac{ \lvert x - x^\prime \rvert}{\theta} + \frac{5}{3} \frac{\prt{x - x^\prime}^2}{\theta^2}} \exp\prt{-\sqrt{5} \frac{\lvert x - x^\prime \rvert}{\theta}}.
\end{equation}
For multidimensional problems, the auto-correlation function is obtained as a tensor product of the unidimensional functions.

The training of a Kriging model is first based on building an \emph{experimental design} which consists of a set of input realizations $\mathcal{X} = \acc{\ve{\chi}^{(i)}, i = 1 \enum n}$ and their corresponding model evaluations $\ve{\mathcal{Y}} = \acc{\mathcal{Y}^{(i)} = \mathcal{M}\prt{\ve{\chi}^{(i)}}, i = 1 \enum n}$. Given these data, a generalized least-square estimate of the weights:
\begin{equation}
\widehat{\ve{\beta}}\prt{\ve{\theta}} = \big( \ve{F}^{T} \ve{R}^{-1} \ve{F}\big)^{-1}\ve{F}^T \ve{R}^{-1}\ve{\mathcal{Y}}
\end{equation}
and the variance estimate:
\begin{equation}
\widehat{\sigma}^2\prt{\ve{\theta}} = \frac{1}{N} \prt{\mathcal{Y} - \ve{F} \widehat{\ve{\beta}}}^T \ve{R}^{-1} \prt{\mathcal{Y} - \ve{F} \widehat{\ve{\beta}}}
\end{equation}
can be derived. In these equations, $ \ve{F} $ is a matrix gathering the regression functions evaluated on the training points, \ie $ F_{ij} = f_j(\ve{\chi}^{(i)})$ and $\ve{R}$ is the auto-correlation matrix defined such that $R_{ij} = R\prt{\ve{\chi^{(i)}}, \ve{\chi^{(j)}}; \ve{\theta}}$.

Once the model is trained, the prediction for any given new point $\ve{x}$ follows a normal distribution, \ie $\widetilde{\mathcal{M}} \sim \mathcal{N}\prt{\mu_{\widetilde{\mathcal{M}}}\prt{\ve{x}}, \sigma^2_{\widetilde{\mathcal{M}}}\prt{\ve{x}}}$ where the mean and variance respectively read:

\begin{subequations}
	\begin{align}
	\mu_{\widetilde{\mathcal{M} }}(\ve{x})  & = f^T(\ve{x})\ve{\beta}+\ve{r}^T(\ve{x})\ve{R}^{-1}\big(\ve{y}-\ve{F}^T\ve{\beta}\big),\label{eq:KrigMean}\\
	\sigma_{\widetilde{\mathcal{M} }}^2(\ve{x})  & = \sigma^2 \Big(1-\ve{r}^T(\ve{x})\ve{R}^{-1}\ve{r}(x) +  \ve{u}^T(\ve{x}) \big(\ve{F}^T \ve{R}^{-1} \ve{F}\big)^{-1} \ve{u}(\ve{x})  \Big),\label{eq:KrigVar}
	\end{align}
\end{subequations}

with $\ve{r}(\ve{x})  = \bra{ R\prt{\ve{x},\ve{\chi}^{(1)}} \enum R\prt{\ve{x},\ve{\chi}^{(n)}}} $ and $ \ve{u}\prt{\ve{x}}  = \ve{F}^T \ve{R}^{-1} \ve{r}\prt{\ve{x}} - \ve{f}\prt{\ve{x}}$.

It remains now to estimate the hyperparameters of the auto-correlation function. Various methods have been proposed to achieve this goal, among which \emph{cross-validation} and \emph{maximum likelihood estimation} \citep{Bachoc2013b}. In this work we consider the latter, which \emph{in fine} consists in solving the following optimization problem \citep{DubourgThesis}:
\begin{equation}
\widehat{\ve{\theta}} = \arg \min_{\ve{\theta} \in n_\theta} \Psi\prt{\ve{\theta}} = \widehat{\sigma}^2 \prt{\ve{\theta}} \lvert \ve{R}\prt{\ve{\theta}}\rvert^{\frac{1}{n}},
\end{equation}
where $\Psi$ is the so-called \emph{reduced likelihood function}, $n_\theta$ is the number of hyperparameters to calibrate and $\lvert \bullet \rvert$ here denotes the determinant operator.

One of the most important aspects of Kriging is its variance (Eq.~\ref{eq:KrigVar})  which can be seen as a local measure of the accuracy of the surrogate model. Typically, such information can be used in active learning when one attempts to build a surrogate model by adaptively refining it so as to ensure sufficient accuracy in some regions of interest. 

Such regions depend on the analyst's aim. In the case when the Kriging model is used to emulate an objective function in an optimization process, the regions of interest are areas of local minima (or maxima). In the case when the analyst is interested in reliability analysis, the Kriging model replaces the limit-state function. The region of interest here corresponds to the vicinity of the limit-state surface. Both cases have been widely studied in the literature under the frameworks of \emph{efficient global optimization} and \emph{efficient global reliability analysis}, respectively \citet{Bichon2008,Bichon2011}. In this paper, both are of interest to us and are briefly described in the sequel.

\subsection{Efficient Global Optimization (EGO)}
Efficient global optimization has been introduced by \citet{Jones1998} as a means to solve optimization problems while replacing the objective function with a Kriging surrogate. The basic idea is to make use of the Kriging variance so as to balance exploitation of areas where the surrogate is minimized and those where its variance is high (due to the lack of data). The algorithm starts by fairly sampling the input space and building an initial Kriging model. Then a \emph{merit function} is used to decide the next point to add in the experimental design so as to bring in the most useful information for the location of the global minimum. Various merit functions have been introduced in the literature. We are interested here in the function used in the contribution of \citet{Jones1998} and originally introduced by \citet{Mockus1974}, the so-called \emph{expected improvement} function, which reads: 
\begin{equation} \label{eq:EIx}
EI(\ve{x})  = (y_{\min}- \Mmean) \Phi \Bigg(  \dfrac{y_{\min}- \Mmean}{ \Mvar}  \Bigg) +  \Mvar \varphi \Bigg(  \dfrac{y_{\min}- \Mmean}{ \Mvar}  \Bigg)
\end{equation}
where $ \varphi $ and $\Phi$ are the standard Gaussian PDF and CDF and $y_{\min}$ is the current known minimum.
This function is made of two complementary parts: the first part relates to the probability of improvement while the second one is proportional to the Kriging variance. By combining these two aspects the expected improvement function achieves its goal of both exploiting and exploring the design space. Hence, the next point to add in the experimental design is chosen as the one that maximizes this function. In the original paper, \citet{Jones1998} uses an optimization algorithm, namely the branch-and-bound algorithm, to locate the global maximum. Here we simply rely on an approximate stochastic (discrete) search. The EGO procedure is then the following:
\begin{enumerate}
	\item Generate a large sample set $N_{\text{EGO}}$ of candidates for enrichment $\mathcal{S} = \acc{\ve{s}^{(1)} \enum \ve{s}^{(N_{\text{EGO}})} }$, where $\ve{s}^{(i)} \in \mathbb{D}$;
	\item Generate an initial experimental design $\mathcal{D} = \acc{\ve{d}^{(1)} \enum \ve{d}^{(m)}}$ and evaluate the corresponding costs $\mathcal{C} = \acc{C_T \prt{\ve{d}^{(1)}} \enum C_T \prt{\ve{d}^{(m)}}}$;
	\item Train a Kriging model $\widetilde{C}_T$ using the experimental design $\acc{\mathcal{D}, \mathcal{C}}$; 
	\item Evaluate the expected improvement function using the candidate set $\mathcal{S}$: \newline $\mathcal{E} = \acc{EI\prt{\ve{s}^{(1)}} \enum EI \prt{\ve{s}^{(N_{\text{EGO}})}}}$ ;
	\item Choose the next best point as the one that maximizes EI on the set $\mathcal{S}$:
	\begin{equation}
	\ve{s}^\ast = \arg \max_{\ve{s} \in \mathcal{S}} \mathcal{E};
	\end{equation}
	\item Check if the convergence criteria are met. If they are, skip to 8, otherwise continue with step 7;
	\item Evaluate $C_T\prt{\ve{s}^\ast}$ and add the couple $\acc{\ve{s}^\ast, C_T\prt{\ve{s}^\ast}}$ to the experimental design $\acc{\mathcal{D}, \mathcal{C}}$. Return to step 3;
	\item End the algorithm.	
\end{enumerate}
For the applications in this paper, $N_{\text{EGO}}$ is set to $10^5$, and points in $\mathcal{S}$ are generated using Latin Hypercube Sampling over the design space. The algorithm stops when $\max_{\mathcal{E}} EI{\prt{\ve{s}}}$ is lower than a given threshold, as suggested in \citet{Bichon2010}. In this paper, a threshold of $10^{-3}$ was found to be adequate for all studied examples.

\subsection{Efficient global reliability analysis (EGRA)}
Even when using EGO, time-variant risk optimization problems may still be computationally intractable. This is due to the necessity of running a full time-variant reliability analysis for each cost evaluation $C_T\prt{\ve{d}}$ that relies on a possibly expensive-to-evaluate limit-state function. The direct approach to cope with this issue is to introduce a meta-modeling at this level, too.  In the most general case, this could involve dynamic problems for which metamodeling is still a challenging issue (See for instance \citet{MaiIJUQ2016} and \citet{MaiSIAMUQ2017}). In this work, we limit our scope to time-variant problems where the limit-state function is characterized by time-independent models (e.g. quasi-static problems). In such a case, a Kriging model $\widetilde{g}$ can be directly used to surrogate the limit-state function, hence further improving the efficiency of the solution.

The reliability counterpart of EGO has been introduced by \citet{Bichon2008} under the name of \emph{efficient global reliability analysis}. It consists in adaptively building a surrogate model so as to ensure accuracy in the vicinity of the limit-state surface. Similarly to EGO, various merit functions, herein known as \emph{learning} functions, have been proposed in the literature. For Kriging this includes, among others, the expected improvement for contour estimation \citep{Ranjan2008}, the deviation number \citep{Echard2011} or the margin probability function \citep{Dubourg2012}. In this work, we consider the so-called \emph{expected feasibility function} proposed by \citet{Bichon2008}, which reads:
\begin{equation} \label{eq:EFF}
\begin{split}
EF(\ve{x})  = \Gmean  \Bigg[ 2 \Phi  \bigg( \frac{\Gmean}{\Gvar} \bigg) - \Phi  \bigg( \frac{-\eps -\Gmean}{\Gvar} \bigg) - \Phi  \bigg( \frac{\eps -\Gmean}{\Gvar} \bigg) \Bigg] \\  - \Gvar \Bigg[  2 \varphi  \bigg( \frac{\Gmean}{\Gvar} \bigg) - \varphi  \bigg( \frac{-\eps -\Gmean}{\Gvar} \bigg) - \varphi  \bigg( \frac{\eps -\Gmean}{\Gvar} \bigg) \Bigg]   \\  + \eps \Bigg[ \Phi \bigg( \frac{\eps - \Gmean}{\Gvar}\bigg) -  \Phi \bigg( \frac{-\eps - \Gmean}{\Gvar}\bigg) \Bigg].
\end{split}
\end{equation}
This function behaves in a similar way as the expected improvement: it takes high values when the evaluated point is close to the limit-state surface and/or when the Kriging variance is high. The next best point to add in the experimental design in order to refine the limit-state surface is therefore the one that maximizes Eq.~\eqref{eq:EFF}. Instead of directly solving this optimization problem, we rely on an approximate discrete procedure as proposed by \citet{Echard2011} under the framework of \emph{active Kriging - Monte Carlo simulation} (AK-MCS) (see also \citet{SchoebiASCE2017}). The algorithm is as follows:
\begin{enumerate}
	\item Generate a large sample set $N_{\text{EGRA}}$ of candidates for enrichment $\mathcal{W} = \acc{\ve{w}^{(1)} \enum \ve{w}^{(N_{\text{EGRA}})} }$. It is worth mentioning here that these samples are drawn in the so-called \emph{augmented space}, a space that encompasses both design and random variables \citep{Kharmanda2002, Au2005, Taflanidis2008, Dubourg2011}. This allows us to build one single global metamodel that can be used for the reliability analysis regardless of the design choice. Details on how such an augmented space is built here can be found in \citet{MoustaphaSMO2016} ;
	\item Generate an initial experimental design $\mathcal{X} = \acc{\ve{x}^{(1)} \enum \ve{x}^{(N_0)}}$ and evaluate the corresponding limit-state responses $\mathcal{G} = \acc{ g\prt{\ve{x}^{(1)}} \enum g \prt{\ve{x}^{(N_0)}}}$;
	\item Train a Kriging model $\widetilde{g}$ using the experimental design $\acc{\mathcal{X}, \mathcal{G}}$; 
	\item Evaluate the expected feasibility function using the candidate set $\mathcal{W}$: \newline $\mathcal{F} = \acc{EF\prt{\ve{w}^{(1)}} \enum EF \prt{\ve{w}^{(N_{\text{EGRA}})}}}$ ;
	\item Choose the next best point as the one that maximizes $EF$ on the set $\mathcal{W}$:
	\begin{equation}
	\ve{w}^\ast = \arg \max_{\ve{w} \in \mathcal{W}} \mathcal{F};
	\end{equation}
	\item Check if the convergence criteria are met. If they are, skip to 8, otherwise continue with step 7;
	\item Evaluate $g\prt{\ve{w}^\ast}$ and add the couple $\acc{\ve{w}^\ast, g\prt{\ve{w}^\ast}}$ to the experimental design $\acc{\mathcal{X}, \mathcal{G}}$. Return to step 3;
	\item End the algorithm.	
\end{enumerate}
In this paper $N_{\text{EGRA}}$ is set to $10^5$ and the points are selected using Latin Hypercube Sampling over the augmented space. Convergence is assumed when $\max_{\mathcal{F}} EFF{\prt{\ve{w}}}$ is lower than a given threshold set to $10^{-3}$.

\end{document}